\documentclass[11pt]{wlscirep}

\usepackage{graphicx}
\usepackage{epstopdf,epsfig}
\usepackage{newtxtext}
\usepackage{newtxmath}
\usepackage{hyperref}
\hypersetup{
    colorlinks = true,
    urlcolor   = blue,
    citecolor  = black,
    linkcolor = blue
}

\newcommand{\RomanNumeralCaps}[1]
\linenumbers

\usepackage{tikz}
\usetikzlibrary{backgrounds,patterns, shapes,calc,arrows,fit,shadows,arrows,positioning,decorations.text,mindmap}

\usepackage[utf8]{inputenc}
\usepackage{overpic}
\usepackage{url}
\usepackage{amssymb}
\usepackage{amsmath}
\usepackage{booktabs}
\usepackage{stmaryrd}
\usepackage{color}
\usepackage{graphicx}
\usepackage{mathtools}
\usepackage{overpic}
\usepackage{subcaption}
\usepackage{mathtools}

\usepackage{algpseudocode}
\usepackage{algorithm}
\usepackage{enumitem}
\usepackage{lipsum}
\usepackage{pifont}
\usepackage{multirow}
\usepackage{cleveref}

\newcommand{\R}{\mathbb{R}}

\newcommand{\dtheta}{\, \mathrm{d} \theta}
\newcommand{\dx}{\, \mathrm{d} x}

\newcommand{\ds}{\, \mathrm{d} s}
\newcommand{\dr}{\, \mathrm{d} r}

\newcommand{\dt}{\, \mathrm{d} t}
\newcommand{\triang}{\mathcal{T}}

\newcommand{\pdel}[1]{\partial_#1}

\newcommand{\cmark}{\ding{51}}%
\newcommand{\xmark}{\ding{55}}%

\newcommand{\pvs}[1]{pvs}

\usepackage{todonotes}

\title{Geometrically reduced modelling of pulsatile flow in perivascular networks}

\author[1, *]{Cécile Daversin-Catty}
\author[1]{Ingeborg G.~Gjerde}
\author[1, 2]{Marie E. Rognes}

\affil[1]{Simula Research Laboratory, Kristian Augusts gate 23, 0164 Oslo, Norway}
\affil[2]{Department of Mathematics, University of Bergen, Norway.}
\affil[*]{cecile@simula.no}

\begin{document}

\begin{abstract}
Flow of cerebrospinal fluid in perivascular spaces is a key mechanism underlying brain transport and clearance. In this paper, we present a mathematical and numerical formalism for reduced models of pulsatile viscous fluid flow in networks of generalized annular cylinders. We apply this framework to study cerebrospinal fluid flow in perivascular spaces induced by pressure differences, cardiac pulse wave-induced vascular wall motion and vasomotion. The reduced models provide approximations of the cross-section average pressure and cross-section flux, both defined over the topologically one-dimensional centerlines of the network geometry. Comparing the full and reduced model predictions, we find that the reduced models capture pulsatile flow characteristics and provide accurate pressure and flux predictions across the range of idealized and image-based scenarios investigated -- at a fraction of the computational cost of the corresponding full models. The framework presented thus provides a robust and effective computational approach for large scale in-silico studies of pulsatile perivascular fluid flow and transport. 
\end{abstract}

\maketitle

\section{Introduction}

Flow of cerebrospinal fluid (CSF) in perivascular spaces (PVSs) is a key transport mechanism in and around the brain~\cite{rennels1985evidence, carare2008solutes, iliff2012paravascular}. A PVS is a space or potential space along or around a blood vessel through which fluid and particles can pass~\cite{wardlaw2020perivascular}. Such spaces appear along blood vessels on the brain surface (surface or pial PVSs) or along blood vessels within the brain parenchyma (parenchymal PVSs). While their shape and structure, and to some extent existence, remain disputed~\cite{zhang1990interrelationships, bedussi2018paravascular, tithof2019hydraulic, min2020surface, wardlaw2020perivascular}, PVSs are typically represented as (elliptic) annular structures or pipes surrounding the blood vessels. As such, surface and parenchymal PVSs form structural networks, dual to and in close interaction with the vascular network, and the surrounding brain tissue and/or subarachnoid space.

Mathematical and computational models are playing an increasingly important role in understanding and predicting PVS flow characteristics~\cite{martinac2019computational}. Theoretical models have quantified the resistance in PVS networks~\cite{faghih2018bulk}, while detailed numerical simulations can predict perivascular fluid velocities and pressures in idealized~\cite{asgari2016glymphatic, diem2017arterial,  rey2018pulsatile, sharp2019dispersion, lloyd2019effects, kedarasetti2020arterial, kedarasetti2020functional} and image-based geometries~\cite{daversin2020mechanisms}. However, computational fluid dynamics simulations rapidly become prohibitively expensive for large, three-dimensional PVS networks. A natural question is therefore whether reduced models can accurately capture PVS flow and transport characteristics and magnitudes. Of particular interest and relevance are geometrically-reduced models for which the computational domain is reduced from an initial three-dimensional representation to a network of topologically one-dimensional branches. Such models have been subject to active research over the last decades in the context of the vasculature, arterial blood flow, and tissue perfusion~\cite{olufsen1999structured, sherwin2003one, d2008coupling, lesinigo2011multiscale, coccarelli2021framework,Kppl2020,Koch2020,vidotto2018, cattaneo2014, Possenti2018, Possenti2021}. For the one-dimensional arterial blood flow models, see e.g.~the seminal work of Olufsen~\cite{olufsen1999structured}, the vasculature is typically represented by a branching network of centerlines, and the model variables are the time-varying cross-section flux and vascular area. The corresponding PVS flow setting has received less attention from the mathematical and numerical community on the other hand. 

In this work, we introduce a geometrically-reduced mathematical model and numerical solution techniques for the time-dependent flow of an incompressible viscous fluid such as CSF in surface PVS networks. The cross-section flux and average pressure are the primary model variables. We consider different computational scenarios including PVS flow induced by a systemic pressure gradient, by cardiac pulse wave-induced movement of the inner vascular wall and by vasomotion in idealized or image-based model geometries. We evaluate the accuracy and efficiency of the reduced models by qualitative and quantitative comparison with the full three-dimensional model analogues. 

The reduced models provide accurate approximations of the cross-section average pressure, cross-section flux and net flow in all geometries considered with relative model discrepancies in the peak flux between $0$ and $35\%$ and in the peak pressure between $0$ and $52\%$. For realistic three-dimensional geometries, the reduced model reduces the computational costs (memory and runtime) by factors of $50-200\times$ with higher factors expected for larger scale networks. 

\section{Methods}

\subsection{PVS geometries (3D and 1D)} 
In general, we consider a perivascular tree-like domain $\Omega$ consisting of a network of branching generalized annular cylinders $\Omega^i$,  with $\Omega \subseteq \cup_{i \in I} \Omega^i$, spatial coordinates $x \in \Omega$ and time $t \geq 0$. The boundary is denoted $\partial \Omega$, with boundary normal $n$. We assume that each generalized annular cylinder $\Omega^i$ has a well-defined and oriented, topologically one-dimensional centerline $\Lambda^i$ with coordinate $s$. We set $\Lambda = \cup_{i \in I} \Lambda^i$. Along $s$, we define the cross-sections $C^i = C^i(s, t)$ of $\Lambda^i$ with area $A^i = A^i(s, t)$. We denote the inner radius of $\Omega^i$ by $R_1^i$ and the outer radius of $\Omega^i$ by $R_2^i$; these radii will in practice vary with $s, t$ and the angular coordinate $\theta$. We denote the set of bifurcation points i.e.~the points at which the centerlines of branches meet by $\mathcal{B}$.

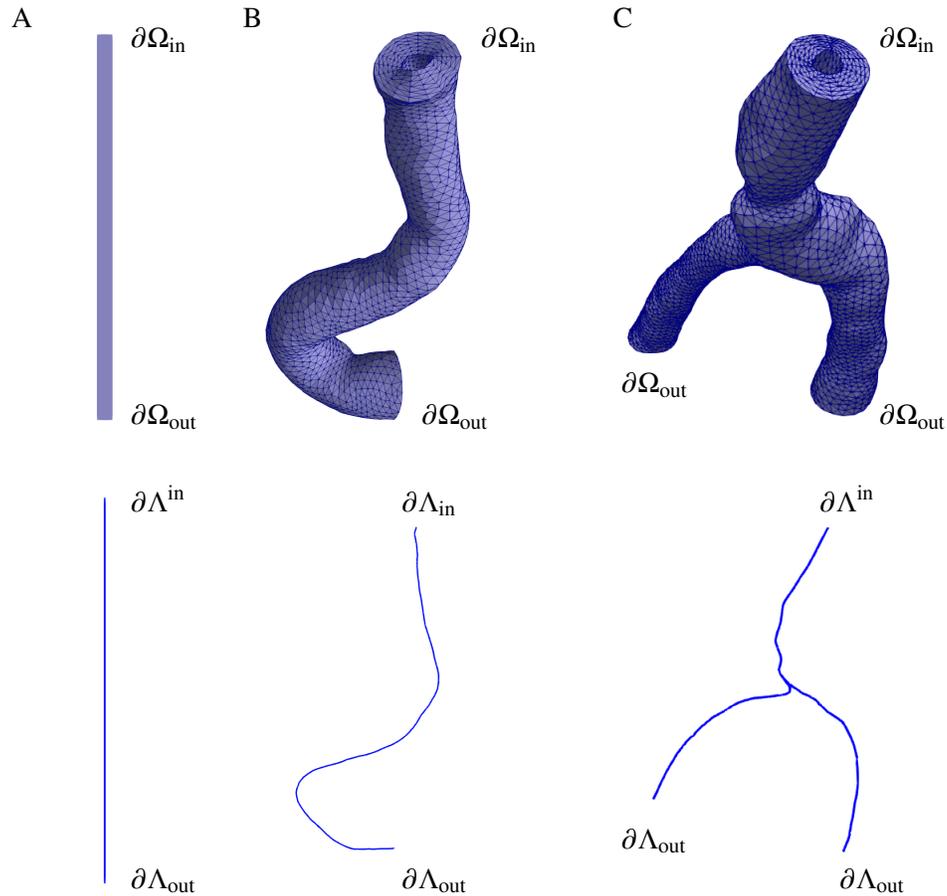
\begin{figure}
    \centering
    \setlength{\tabcolsep}{28pt}
    \begin{tabular}{ccc}
        \begin{overpic}[height=0.3\textwidth]{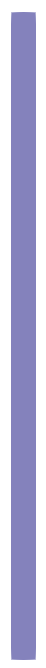}
        % A : Straight line - 2D
        \put(-20,100){A}
        \put(10,0){$\displaystyle\partial\Omega_{\mathrm{out}}$}
        \put(10,95){$\displaystyle\partial\Omega_{\mathrm{in}}$}
        \end{overpic}
        &
        % B : C0092 - 3D
        \begin{overpic}[height=0.3\textwidth]{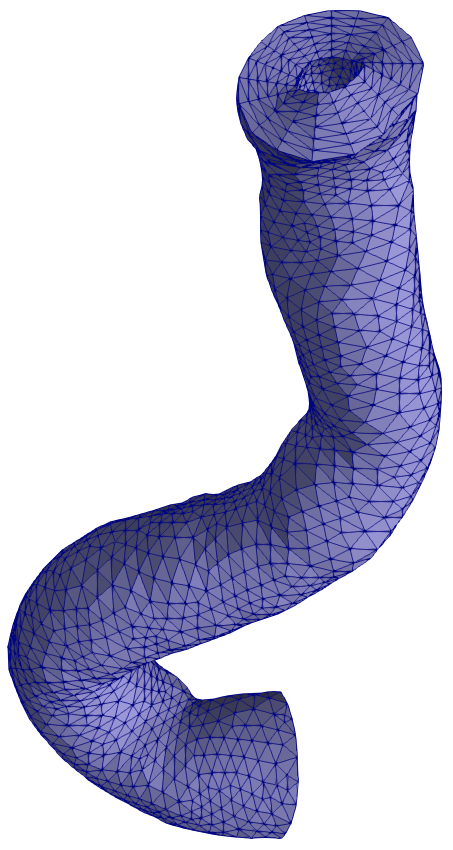}
        \put(-5,100){B}
        \put(40,0){$\displaystyle\partial\Omega_{\mathrm{out}}$}
        \put(55,95){$\displaystyle\partial\Omega_{\mathrm{in}}$}
        \end{overpic}
        &
        % C : C0075 - 3D
        \begin{overpic}[height=0.3\textwidth]{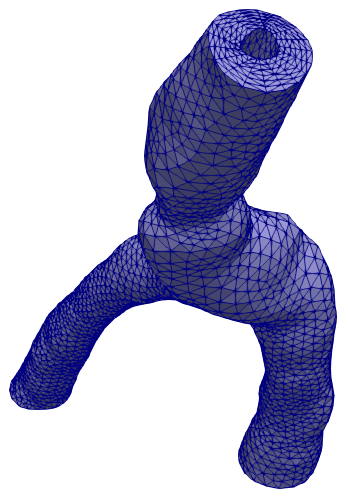}
        \put(-2,100){C}
        \put(0,8){$\displaystyle\partial\Omega_{\mathrm{out}}$}
        \put(65,0){$\displaystyle\partial\Omega_{\mathrm{out}}$}
        \put(65,95){$\displaystyle\partial\Omega_{\mathrm{in}}$}
        \end{overpic}
        \\[0.75cm]
        % A : Straight line - 1D
        \begin{overpic}[height=0.3\textwidth]{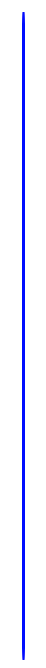}
        %\put(-20,100){\small \textbf{A-2)}}
        \put(10,0){$\displaystyle \partial \Lambda_{\mathrm{out}}$}
        \put(10,95){$\displaystyle\partial \Lambda^{\mathrm{in}}$}
        \end{overpic}
        &
        % B : C0092 - 1D
        \begin{overpic}[height=0.3\textwidth]{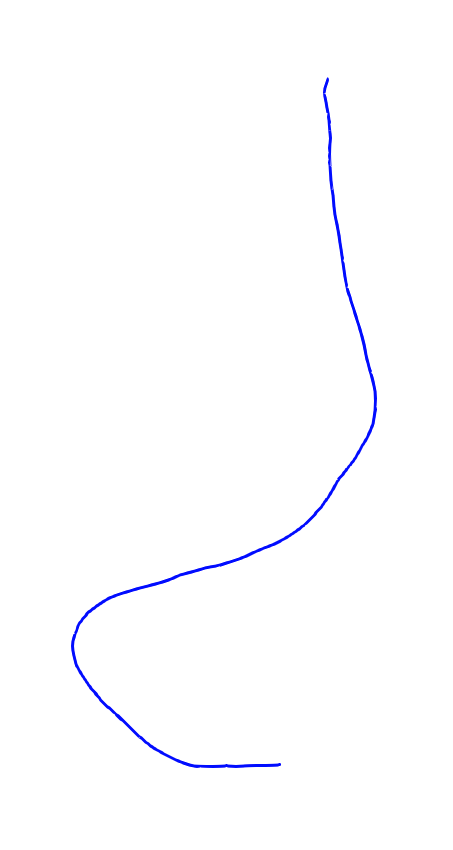}
        %\put(-5,100){\small \textbf{B-2)}}
        \put(35,0){$\displaystyle \partial \Lambda_{\mathrm{out}}$}
        \put(35,95){$\displaystyle\partial\Lambda_{\mathrm{in}}$}
        \end{overpic}
        &
        % C : C0075 - 1D
        \begin{overpic}[height=0.3\textwidth]{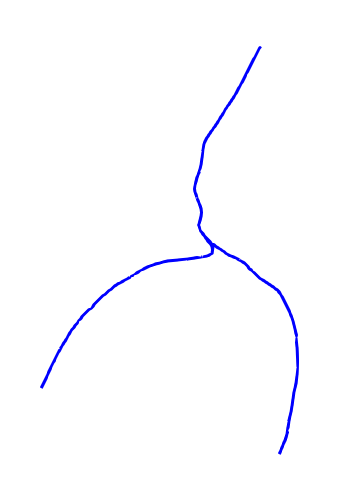}
        %\put(-2,100){\small \textbf{C-2)}}
        \put(0,10){$\displaystyle \partial \Lambda_{\mathrm{out}}$}
        \put(55,0){$\displaystyle \partial \Lambda_{\mathrm{out}}$}
        \put(50,95){$\displaystyle\partial \Lambda^{\mathrm{in}}$}
        \end{overpic}
        \\
    \end{tabular}
    % C0092 : MCA (middle cerebral artery) - age:62, sex:M
    \caption{Overview of the full three-dimensional and topologically one-dimensional reduced model domains. The idealized geometry A (the \emph{axisymmetric PVS}) is a single $1$ mm long axisymmetric annular cylinder represented by its two-dimensional angular cross-section. Geometry B (the \emph{image-based PVS}) is generated from a cerebral artery segment (Aneurisk dataset repository, case id C0092) and represents a realistic perivascular space without bifurcation. Geometry C (the \emph{bifurcating image-based PVS}) is generated from a middle cerebral artery (MCA M1--M2) segment (Aneurisk dataset repository, case id C0075) and represents a realistic perivascular space including a bifurcation.}
    \label{fig:domains}
\end{figure}

We introduce three specific geometries of increasing complexity: from an axisymmetric cylinder (A) to an image-based perivascular geometry without any bifurcations (B) and one with a bifurcation (C) (Figure \ref{fig:domains} and Table \ref{tab:geometry-data}). Three-dimensional PVS flow in geometries A and C have been studied previously~\cite{daversin2020mechanisms} and will be used for comparison. In each of these geometries, the PVS domain is defined by creating a generalized annular cylinder surrounding the vascular segment with the vascular wall as the inner surface of the PVS. The width of the PVS is set proportional to the blood vessel diameter (by factor of $0.95$) and scaled (to a mouse scale)~\cite{daversin2020mechanisms, mestre2018flow}. We define as PVS inlets and outlets ($\partial \Omega_{\rm in}$ and $\partial \Omega_{\rm out}$) the PVS ends surrounding the vascular inlets and outlets, respectively, noting however that fluid may flow both in and out of both the inlet and outlets. We denote the inner PVS wall (boundary) by $\partial \Omega_{\rm inner}$ and outer wall by $\partial \Omega_{\rm outer}$.

\begin{table}
    \centering
    \begin{tabular}{crrr|rrr|rr}
    \toprule
    Domain & 
    $L$ (mm) & 
    $D_a$ (mm) &
    $D_{\rm pvs}$ (mm) &
    \multicolumn{3}{c}{Mesh (Full)} & 
    \multicolumn{2}{c}{Mesh (Reduced)} \\
    & & & & cells & vertices & $h_{\min}$ (mm) & vertices & $h_{\min}$ (mm) \\
    \midrule
    A & 1 & 0.04 & 0.06 & 1920 & 1053 & $1.3 \times 10^{-2}$ & 65 & $1.6 \times 10^{-2}$ \\
    B & $\approx$1 & 0.036--0.047 & 0.035--0.044 & 63144 & 12404 & $9.4 \times 10^{-3}$ & 356 & $2.8 \times 10^{-4}$ \\
    C & $\approx$1 & 0.024--0.046 & 0.023--0.044 & 88074 & 17318 & $6.4 \times 10^{-3}$ & 249 & $9.9 \times 10^{-5}$ \\
    \bottomrule
    \end{tabular}
    \caption{Geometrical or numerical PVS domain characteristics for domains A, B, C. $L$ denotes an approximate domain length, $D_{a} = 2 R_1$ is the range of the arterial diameters, $D_{\rm pvs}$ indicates the range of widths of the perivascular space ($D_{\rm pvs} = R_2 - R_1$, so that $R_2 = 2.95 R_1$) cells and vertices indicate the number of mesh cells and mesh vertices respectively for the full (2D or 3D) model and reduced models, and $h_{\max}$ denotes the maximal mesh cell size for each mesh. The vertices for the one-dimensional geometries are uniformly spaced in the interior of the domain.}
    \label{tab:geometry-data}
\end{table}

The 3D PVS construction and the 1D centerline extraction are performed using \textit{PVS-meshing-tools}~\cite{pvs-meshing-tools}, largely based on VMTK~\cite{VMTKAntiga2008AnIM}. The extracted centerline comes with underlying data including the branch lengths and vessel radii. The centerline radius refers to the radius of the maximal inscribed circle of the vessel cross-sections. The meshing of both 3D and 1D PVS domains is performed within \textit{PVS-meshing-tools}~\cite{pvs-meshing-tools} using meshio~\cite{meshio} and GMSH~\cite{GMSH}. The centerline meshes consist of topologically one-dimensional intervals embedded in three dimensions. The bifurcation points $b \in \mathcal{B} \subset \Omega$ are explicitly labeled within each centerline mesh. Each branch is also separately tagged and given a consistent orientation. This procedure allows for the identification of bifurcation points as the outlet of one (parent) centerline and the inlet of other (daughter) centerlines, and a split of the full perivascular network into oriented mesh branches. 

\subsection{Stokes flow in a deforming perivascular domain}

Flow of CSF in surface PVSs is reported to be laminar, with low Reynolds numbers ($10^{-4}-10^{-2}$) and moderate P\'eclet numbers ($10^{2}-10^{4}$), a mean flow speed of up to 60 $\mu$m/s, and parabolic flow profiles~\cite{mestre2018flow}. We therefore model the flow of an incompressible, viscous fluid flowing at low Reynolds and Womersley numbers via the time-dependent Stokes equations over a time-dependent domain $\Omega = \Omega(t)$ representing the PVS. The fluid velocity $v = v(x, t)$ for $x \in \Omega(t)$ at time $t$ and the CSF pressure $p = p(x, t)$ then solve the following system of time-dependent partial differential equations (PDEs)~\cite{san2009convergence, daversin2020mechanisms}:
\begin{subequations}
    \begin{align}
        \rho \pdel{t}v - \mu\nabla^2 v + \nabla p = 0 \quad \text{ in } \Omega(t), \\
        \nabla \cdot v = 0 \quad \text{ in } \Omega(t),
    \end{align}
    \label{eq:stokes}%
\end{subequations}  
where $\rho$ is the fluid density and $\mu$ is the dynamic fluid
viscosity. To model CSF at body temperature, we set the fluid density to $\rho = 10^{3}$ kg/m$^3$ and the dynamic viscosity to $\mu = 0.697\times 10^{-3}$ Pa~s. As in our previous full models of perivascular flow~\cite{daversin2020mechanisms}, the initial PVS mesh defines the reference domain $\Omega(0)$, and we assume that $\Omega(t)$ at time $t > 0$ is given by a deformation $d$ of the reference domain: $\Omega(0) \mapsto \Omega(t)$ with $x = d(X, t)$, $X \in \Omega(0)$, $x \in \Omega(t)$. We denote the domain velocity associated with $d$ by $w$ (thus $\dot d = w$). 

\subsection{Boundary conditions, initial conditions and periodicity}
At the PVS ends, we prescribe a traction condition corresponding to a known, applied pressure $\tilde p = \tilde p(x, t)$: 
\begin{equation}
   \sigma_n \equiv (\mu \nabla u - p I) \cdot n = - \tilde{p} n \qquad\text{ on } \partial \Omega_{\text{in}} \text{ and } \partial \Omega_{\text{out}}. 
\label{eq:bcs}
\end{equation}We either prescribe (i) zero pressure at both ends $\tilde{p} = 0$, or (ii) a constant-in-time pressure gradient $\Delta \tilde{p} > 0$ by setting $\tilde p_{\rm in} = L_{\rm out} \Delta \tilde{p}$ at the inlet, letting $\tilde{p}_{\rm out} = 0$ at the outlet furthest from the inlet with distance $L_{\rm out}$, and setting $\tilde{p}_{\rm out}$ at any other outlets such that the average pressure gradient over each branch path $(\tilde{p}_i - \tilde{p}_{\rm out})/L_{\rm out}$ is constant and equal to the prescribed pressure gradient $\Delta \tilde{p}$ mmHg/m. This static pressure difference can represent e.g.~a hydrostatic pressure difference, a venous pressure differential, or some other systemic pressure difference. 

On the inner and outer PVS walls (along the length of the PVS), we set the fluid velocity $v$ to match a known, prescribed domain velocity $w = w(x, t)$. For the inner PVS wall, we either (i) consider a rigid wall and set $v = w = 0$, or (ii) impose a pulsating wall displacement:
\begin{equation}
    d |_{\partial \Omega_{\rm inner}}(X, t) =  A(X, t) \, n , \label{eq:wall-disp}
\end{equation}
with reference to the initial (fixed) mesh with coordinates $X$ and prescribe $v = w = \dot d$. To represent \emph{wall motion induced by the cardiac pulse wave}, we let the amplitude $A$ be defined by the juxtaposition of an experimentally-observed wall motion time series~\cite{mestre2018flow} either applied uniformly along the length of the PVS or as a travelling wave along the PVS length with wave speed $c$ = 1 m/s and frequency 10 Hz. We refer to~\cite{daversin2020mechanisms} for the detailed description. To represent \emph{wall motion due to vasomotion}, we consider a similar set-up but with a travelling sinusoidal wave in time with a frequency of $0.1$ Hz and wave length $\lambda$ 8 mm~\cite{aldea2019cerebrovascular}, and an amplitude $A$ of $7.5\%$ of the initial inner radius $R_1$. We note that for all models, the wall moves in the normal (radial) direction only. For the outer PVS wall $\partial \Omega_{\rm outer}$, we set $v = w = 0$. 

The system starts at rest with $v = w = 0$ at $t = 0$. The system reaches the periodic steady state nearly immediately, and we report results starting from the first cycle. 

\subsection{Model reduction assumptions}

We define a reduced, topologically one-dimensional, model approximation of the full PVS flow model (\eqref{eq:stokes} with the given boundary and initial conditions) under the following stipulations~\cite{gjerde2021analysis}. For each branch $\Omega^i(t)$ with centerline $\Lambda^i$ and local coordinate system $(s, r, \theta)$, where $s$ represents the path length (or axial coordinate), $r$ is the radial coordinate and $\theta$ is the angular coordinate, we suppose that:
\begin{description}{\leftmargin=-2em}
    \item[(I) ] \textit{Axial symmetry}. Fields and input parameters are independent of the angular coordinate $\theta$;
    \item[(II) ] \textit{Radial displacements.} Boundaries displace in the radial direction only;
    \item[(III) ] \textit{Fixed centerline.} The centerline $\Lambda$ is fixed in time and defines the axial direction; 
    \item[(IV) ] \textit{Constant cross-section pressure.} The pressure field is independent of the angular and radial coordinates i.e.~$p = p(s, t)$;
    \item[(V) ] \textit{Axial velocity profile} The axial velocity $v_s$, i.e.~the velocity component in the axial direction can be decomposed in the form
    \begin{equation}
        v \cdot s = v_s = v_s(s, r, t) = \hat{v}(s,t) v_{\rm vp}(r),
    \end{equation}  
    where $v_{\rm vp}$ is a given velocity profile varying radially only, $\hat{v}$ is to be determined. 
\end{description}
For the velocity profile $v_{\rm vp}$, we here choose a normalized annular Poiseuille flow:
\begin{align}
v_{\rm vp}(r) = \frac{v_{\text{poise}}(r)}{v_{\text{poise}}(\frac{R_1+R_2}{2})},\quad v_{\text{poise}}(r)= \left(  1-\frac{r^2}{R_1^2} +\frac{R_2^2-R_1^2}{R_1^2 \ln\left( R_2/R_1 \right)}\ln(r/R_1)  \right).
\label{eq:velo}
\end{align}
This velocity profile is parabolic in $r$ (as for Poiseuille flow in a cylinder) with a logarithmic correction that accounts for the annulus.

In particular, the domain velocity $w$ is assumed independent of the angular coordinate $\theta$. Note that we do not assume other velocity components (than the axial) to necessarily be zero. We emphasize that these assumptions will in general not be satisfied by realistic geometries and flows. Thus, the reduced model defines a model approximation associated with a certain modelling error. 

\subsection{Reduced model equations}

Under the assumptions \textbf{(I)}-\textbf{(V)}, the full PVS flow model can be reduced to the following system of time-dependent differential equations~\cite{gjerde2021analysis}: find the cross-section flux $\hat{q} = \hat{q}(s, t)$ and the cross-section average pressure $\hat{p} = \hat{p}(s, t)$ such that for each centerline $\Lambda^i$ (denoting $\hat{q}|_{\Lambda^i} = \hat{q}^i$ and $\hat{p}|_{\Lambda^i} = \hat{p}^i)$:
\begin{subequations}
    \begin{align}
        \frac{\rho}{A^i} \pdel{t} \hat{q}^i -     \frac{\mu}{A^i} \partial_{ss} \hat{q}^i + \mu \frac{\alpha^i}{A^i}  \hat{q}^i  + \pdel{s}\hat{p}^i  &=0 && \text{ on } \Lambda^i, \label{eq:1d-pvs-1}\\
        \pdel{s} \hat{q}^i &= \hat{f}^i && \text{ on } \Lambda^i, \label{eq:1d-pvs-2}
    \end{align}
    \label{eq:1d-pvs}
\end{subequations} 
hold. 
\begin{align}
    \hat{f}^i(s) \equiv 2 \pi R_1^i(s, t) w(R_1, s, t) \cdot n|_{\partial \Omega_{\rm inner}} + 2 \pi R_2^i(s, t) w(R_2, s, t) \cdot n|_{\partial \Omega_{\rm outer}}.
\end{align}
Moreover, $A^i = A^i(s, t)$ denotes the cross-section area, while $\hat{\alpha}^i = \hat{\alpha}^i(s, t)$ is a lumped flow parameter that depends on the domain geometry and the choice of velocity profile $v_{\rm vp}$:
\begin{align}
    \alpha^i(s, t) \equiv \frac{1}{A \bar{\bar{v}}_{\rm vp}(s)} \left (
    2 \pi R_1^i(s, t) \, \partial_r v_{\rm vp}(R_1^i (s, t)) - 2 \pi R_2^i(s, t) \, \partial_r v_{\rm vp}(R_2^i(s, t)) \right ),
    \label{eq:alpha}
\end{align}
and where $\bar{\bar{v}}_{\rm vp}$ is the velocity profile integrated over each cross-section:
\begin{equation}
    \bar{\bar{v}}_{\rm vp} \equiv \int_{C(s)} v_{\rm vp} \, r \dr \dtheta.
\end{equation}
We also define the (one-dimensional) normal stress induced by $\hat{q}$ and $\hat{p}$:
\begin{equation}
    \hat{\sigma} \equiv \frac{\mu}{A} \pdel{s} \hat{q}-\hat{p},
    \label{eq:1d-stress}
\end{equation} 
which corresponds to an average of the axial ($s$-)component of the normal stress in \eqref{eq:bcs} over each cross-section.

At the bifurcation points $b \in \mathcal{B} \subset \Omega$, we impose the following two conditions representing conservation of flux and continuity of normal stress, respectively:
\begin{align}
    \label{eq:conservation:flux} 
    \hat{q}^p(s^p) &= \hat{q}^{d_1}(s^{d_1}) + \hat{q}^{d_2}(s^{d_2}), \\
    \label{eq:continuity:stress}
    \hat{\sigma}^p(s^p) &= \hat{\sigma}^{d_1}(s^{d_1}) = \hat{\sigma}^{d_2}(s^{d_2}),
\end{align}
where $\Lambda^p$ and $\Lambda^{d_1}$, $\Lambda^{d_2}$ represent the centerlines of the parent and two daughter branches, respectively, associated with the bifurcation point $b$ and $s^{\cdot} = \iota^{\cdot}(b)$ where $\iota^{\cdot}$ denotes the map from three-dimensional bifurcation point to the one-dimensional centerline coordinate for each branch $\Omega^{\cdot}$. 

The system~\eqref{eq:1d-pvs} defines a set of equations for each branch centerline $\Lambda_i$ and is closed by the bifurcation conditions~\eqref{eq:conservation:flux}--\eqref{eq:continuity:stress}, together with boundary conditions at the PVS inlet and outlets, as well as initial conditions for the cross-section flux. Specifically, in place of the traction condition~\eqref{eq:bcs}, we prescribe the corresponding pressure difference for the (average) normal stress $\hat{\sigma}$ cf.~\eqref{eq:1d-stress}. In this manner, the (one-dimensional) solutions $\hat{q}^i$ and $\hat{p}^i$ of the reduced model~\eqref{eq:1d-pvs} define approximations of the (three-dimensional) axial flux and pressure solving~\eqref{eq:stokes} integrated or averaged over each cross-section:
\begin{align*}
\hat{q}^i(s) &\approx \int_{C^i(s)} v_s (s, r, t) \, r \dr \dtheta \equiv A^i(s) \, \bar{\bar{q}}^i(s) , \\
\hat{p}^i(s) &\approx \frac{1}{A^i(s)} \int_{C^i(s)} 
p (s, r, t)\, r  \dr \dtheta. 
\end{align*}
The factor $r$ originates from integrating in cylindrical coordinates. We note that the wall velocity $w$, which defines a boundary condition for the full PVS model~\eqref{eq:stokes}, enters as a body force in the reduced model~\eqref{eq:1d-pvs}. 

\subsection{Numerical solution and software}

We solve the full PVS equations~\eqref{eq:stokes} via a previously developed and verified arbitrary Lagrangian-Eulerian (ALE) formulation and finite element discretization~\cite{daversin2020mechanisms}. This solver builds on the standard FEniCS finite element software suite~\cite{AlnaesBlechta2015a}, and is openly available~\cite{mechanisms-behind-pvs-flow-zenodo}.

To compute numerical solutions to the reduced model \eqref{eq:1d-pvs}, we consider a first-order implicit Euler scheme in time and a higher-order finite element method in space~\cite{gjerde2021analysis}. The finite element mesh $\triang$ of the centerline $\Lambda$ is composed of mesh segments $\triang^i$, one for each centerline branch $\Lambda^i$. Each mesh segment is a mesh consisting of intervals embedded in $\R^3$. We label the set of bifurcation points $\mathcal{B}$, inlet points $\mathcal{I}$ and outlet points $\mathcal{O}$,  and define the following finite element spaces:
\begin{itemize}
    \item The flux space $V_h$ is the space of continuous piecewise quadratics over $\triang^i$ for each $i$. 
    \item The (average) pressure space $Q_h$ is the space of continuous piecewise linears on $\triang$.
    \item The Lagrange multiplier space $R_h = \R^B$ where $B$ is the number of bifurcation points.
\end{itemize}
The flux is thus solved on each mesh segment representing the PVS network branches and may be discontinuous across bifurcations. We impose the flux conservation condition~\eqref{eq:conservation:flux} weakly using a Lagrange multiplier formulation. The pressure is solved on the whole mesh and is continuous at bifurcations by construction. 

For each discrete time $t^k$, given $\hat{q}_h^{k-1}$ at the previous time $t^{k-1}$ and time step $\Delta t = t^k - t^{k-1}$, we solve for the approximate cross-section flux $\hat{q}^k_h \in V_h$, average pressure $\hat{p}^k_h \in Q_h$ and a Lagrange multiplier (corresponding to the normal stress~\eqref{eq:1d-stress} at the bifurcation points) $\lambda^k_h \in R_h$ solving
\begin{equation}
    \label{eq:variational}
    a((\hat{q}^k_h, \hat{p}^k_h, \lambda^k_h), (\psi, \phi, \xi)) = 
    L^k((\psi, \phi, \xi)),
\end{equation}
for all finite element test functions $\psi \in V_h$, $\phi \in Q_h$, and $\xi \in R_h$. The left-hand side bilinear form $a$ is defined by:
\begin{multline}
a((q, p, \lambda), (\psi, \phi, \xi)) = \\
\sum_{i \in I} \int_{\Lambda^i} 
\frac{1}{A^i} \left (\rho + \Delta t \mu \alpha^i \right ) q^i \psi^i 
+ \frac{\Delta t \mu}{A^i} \partial_s q^i \partial_s \psi^i 
+ \partial_s q^i \phi^i 
- \Delta t  \partial_s \psi^i p^i
\ds 
+ \sum_{b \in \mathcal{B}} \lambda^b [\psi]^b + \xi^b [q]^b,
  \end{multline}
where $\lambda^b$ (or $\xi^b$) is simply the entry of the vector $\lambda$ (or $\xi$) corresponding to bifurcation point $b$, and we define the natural jump:
\begin{equation}
    [\psi]^b = \psi^p (b) 
     - \psi^{d_1} (b) 
     - \psi^{d_2} (b).
\end{equation}
The right-hand side linear form $L$ is: 
\begin{equation}
    L^k(\psi, \phi, \xi) = 
    \sum_{i \in I} \int_{\Lambda^i} 
    \frac{\rho}{A^i} \hat{q}^{k-1, i}_h \psi^i + f^i \phi^i \ds 
    - \sum_{x \in \mathcal{I}}
    \Delta t \tilde{p}_{\rm in} (x) \psi^{i_I} (x) 
    + \sum_{x \in \mathcal{O}}
    \Delta t \tilde{p}_{\rm out} (x) \psi^{i_O} (x),
\end{equation}
where the superscript $i_I$ ($i_O$) in the inlet (outlet) terms above refers to the unique centerline branch associated with the inlet (outlet) points. 

The numerical solver for the reduced model was implemented in the well-established FEniCS Project finite element software~\cite{AlnaesBlechta2015a}. The solver, and in particular the definition of the partially continuous flux space, builds on mixed-domain features~\cite{daversincatty2019abstractions} and relies on the latest development version of FEniCS.  

\subsection{Overview of computational models, output functionals and model error measures}
An overview of the six computational models considered is given in Table~\ref{tab:models-parameters}. Each model is labeled with reference to its domain (A, B, or C) followed by a number indicating the driving forces included: (1) a given pressure drop, (2) wall movement due to cardiac pulsations and (3) wall movement due to vasomotion. For each model, we consider the full three-dimensional version as well as the reduced model. 

To compare the solutions from the full and reduced models, we consider the following quantities of interest. For each domain, we define a set of cross-sections as follows. For domain A, we define the left-most end as the inlet ($s = 0$) and define an upper cross-section. For domain B, we consider the inlet and outlet ends of the PVS, as well as upper and lower cross-sections. For domain C, we consider the inlet at $s = 0$, and the two outlets, as well as three additional cross sections near the inlet, on the largest daughter branch relatively close to the bifurcation, and near the outlet of the other daughter branch. 

For each cross section $C$, we compare a numerical approximation $\bar{\bar{p}}_h$ of the average of the full pressure solution:
\begin{equation}
    \bar{\bar{p}}_h (t) = \frac{1}{|C|} \sum_{k} w_k p_h(x_k, t) 
\end{equation}
for a quadrature scheme with points $x_k$ and weights $w_k$ defined over $C$ and an approximation $|C|$ of the cross-section area~\cite{gjerde2021analysis}. The averaging is implemented by using the Frenet frame associated with $\Lambda$ to map from an annular cylinder in a reference domain onto the cross-section, similar to the implementation of the averaging operator in fenics$\textunderscore$ii \cite{Kuchta2020}.

Similarly, we compute a numerical flux approximation $\bar{\bar{q}}_h$ by the same numerical integration of the axial velocity over the cross-section $C$. 
%We then evaluate the relative model discrepancies in the flux and average %pressure at each cross-section $C$ by
%\begin{equation}
%    E_{C, p}(t) = |\bar{\bar{p}}_h (t) - \hat{p}_h (t)| / 
%    | \bar{\bar{p}}_h (t)|, \quad
%    E_{C, q}(t) = | \bar{\bar{q}}_h (t)  - \hat{q}_h (t) | /
%    | \bar{\bar{q}}_h (t)|. 
%\end{equation}
We define the total relative model discrepancy $E_q$ in the flux by
\begin{equation}
   E_q (T) = \| \bar{\bar{p}}_h (T) - \hat{p}_h(T) \|_{L^2(\Lambda)}
   / \| \bar{\bar{p}}_h (T) \|_{L^2(\Lambda)}
   \label{eq:modelerror}
\end{equation}
and similarly for the pressure $E_p(t)$.

Finally, we define the net flux $Q$ associated with the velocity $v = v(x, t)$ as:
\begin{equation}
    Q = \int_{0}^{T} \int_{\partial \Omega_{\rm in}} v \cdot n \dx \dt,
\end{equation}
and the corresponding quantity associated with the flux $\hat{q} = \hat{q}(s)$:     
\begin{equation}
    Q = \int_{0}^{T} \sum_{x \in \mathcal{I}} \hat{q}(x) \dt
\end{equation}
where the integration in time is over one period $[0, T]$.      
      
\begin{table}
    \newcommand{\mcmark}{\multirow{2}{0.02\textwidth}{\cmark}}
    \newcommand{\mxmark}{\multirow{2}{0.02\textwidth}{\xmark}}
    \centering
    % model | boundary conditions | parameters | Assumptions (1-5)
    \begin{tabular}{lllllllll}
    \toprule
          \multirow{2}{0.15\textwidth}{} & 
          \multirow{2}{0.1\textwidth}{Domain} &
          \multirow{2}{0.22\textwidth}{Pressure gradient \\ $\Delta \tilde{p}$ [Pa/mm]} & 
          \multirow{2}{0.22\textwidth}{Wall motion pattern} & 
          \multicolumn{5}{c}{Model assumptions} \\[0.1cm]
          & & & & (\textbf{I}) & (\textbf{II}) & (\textbf{III}) & (\textbf{IV}) & (\textbf{V}) \\
         \midrule
         %% A1
         \multirow{2}{0.15\textwidth}{Model A1} & 
         \multirow{2}{0.1\textwidth}{A} &
         \multirow{2}{0.22\textwidth}{0.1995}& 
         \multirow{2}{0.22\textwidth}{None}&
         \mcmark & \mcmark & \mcmark & \mcmark & \mcmark \\
         {} & 
         {} & 
         {} & 
         & & & & \\
         %% A2
        \multirow{2}{0.15\textwidth}{Model A2} & 
        \multirow{2}{0.1\textwidth}{A} &
         \multirow{2}{0.22\textwidth}{0.0} & 
         \multirow{2}{0.22\textwidth}{Cardiac pulsations (uniform)} & 
         \mcmark & \mcmark & \mcmark & \mxmark & \mxmark \\
         {} & 
         {} & 
         {} & 
         & & & & \\
         %% B1
         \multirow{2}{0.15\textwidth}{Model B1} & 
         \multirow{2}{0.1\textwidth}{B} &
         \multirow{2}{0.22\textwidth}{0.1995} & 
         \multirow{2}{0.22\textwidth}{None} & 
         \mxmark & \mcmark & \mcmark & \mxmark & \mxmark \\
         {} & 
         {} & 
         {} & 
         & & & & \\
         %\midrule      
         %% B2
        \multirow{2}{0.15\textwidth}{Model B2} & 
        \multirow{2}{0.1\textwidth}{B} &
         \multirow{2}{0.22\textwidth}{0.0} & 
         \multirow{2}{0.22\textwidth}{Cardiac pulsations (travelling)} & 
         \mxmark & \mcmark & \mcmark & \mxmark & \mxmark \\
         {} & 
         {} & 
         {} & 
         & & & & \\
         %% B3
        \multirow{2}{0.15\textwidth}{Model B3} & 
        \multirow{2}{0.1\textwidth}{B} &
         \multirow{2}{0.22\textwidth}{0.0} & 
         \multirow{2}{0.22\textwidth}{Vasomotion (travelling)} & 
         \mxmark & \mcmark & \mcmark & \mxmark & \mxmark \\
         {} & 
         {} & 
         {} & 
         & & & & \\
         %% C
         \multirow{2}{0.15\textwidth}{Model C12} & 
         \multirow{2}{0.1\textwidth}{C} &
         \multirow{2}{0.22\textwidth}{0.1995} & 
         \multirow{2}{0.22\textwidth}{Cardiac pulsations (travelling)} & 
         \mxmark & \mcmark & \mcmark & \mxmark & \mxmark \\
         {} & 
         {} & 
         {} & 
         & & & & \\
         \bottomrule
    \end{tabular}
    \caption{Overview of computational models parameterized by domain, prescribed pressure gradient $\Delta \tilde{p}$ and wall motion pattern (see Methods). Wall pulsations are applied uniformly in space (uniform) or as a travelling wave in space (travelling). Each of the models satisfy some of the reduced model assumptions (I)-(V), but only Model A1 satisfies all.}
    \label{tab:models-parameters}
\end{table}

\section{Results}

The prescribed pressure gradient and the pulsating PVS walls each induce pressure gradients and fluid flow in the different PVS geometries. For each of the models (Table~\ref{tab:models-parameters}), we compare the simulation results from the full PVS equations~\eqref{eq:stokes} defined over the three-dimensional model domains and the reduced system~\eqref{eq:1d-pvs} defined over the topologically one-dimensional domains, quantify the discrepancies between the models and the computational costs.

\begin{figure}
    \centering
    \begin{subfigure}{0.1\textwidth}
    \includegraphics[width=0.8\textwidth]{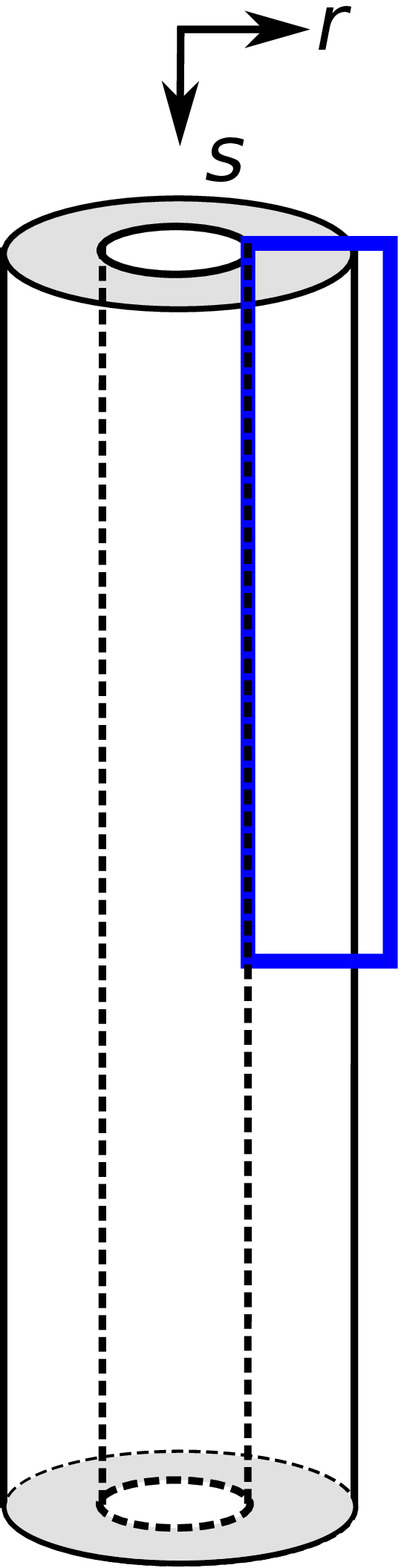}
    \end{subfigure}
    \begin{subfigure}{0.84\textwidth}
    \centering
    \includegraphics[width=0.86\textwidth]{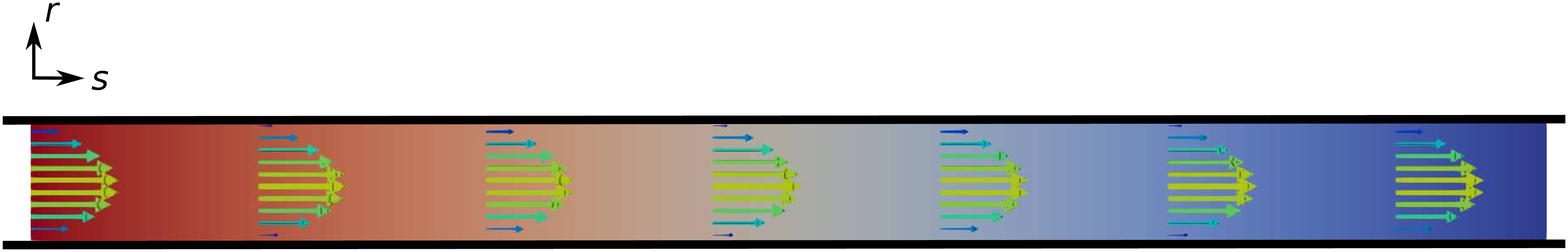} \\
    \vspace{3em}
    \includegraphics[width=0.86\textwidth]{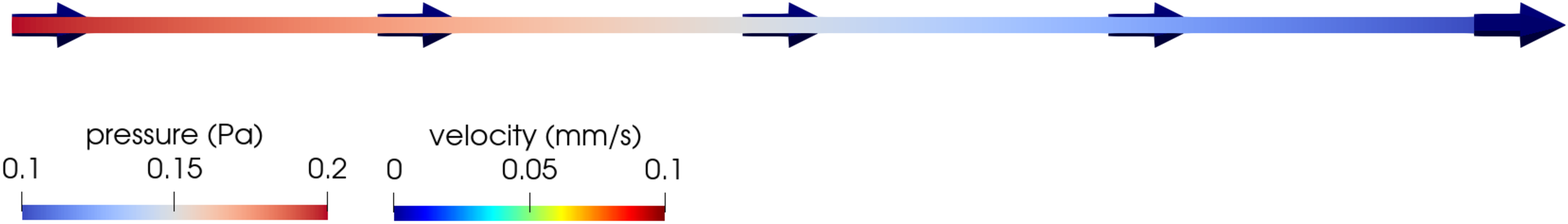} \\
    \vspace{1em}
    \subcaption{}
    \label{fig:A1}
    \end{subfigure}
    \begin{subfigure}{0.84\textwidth}
    \centering
    \includegraphics[width=1.0\textwidth]{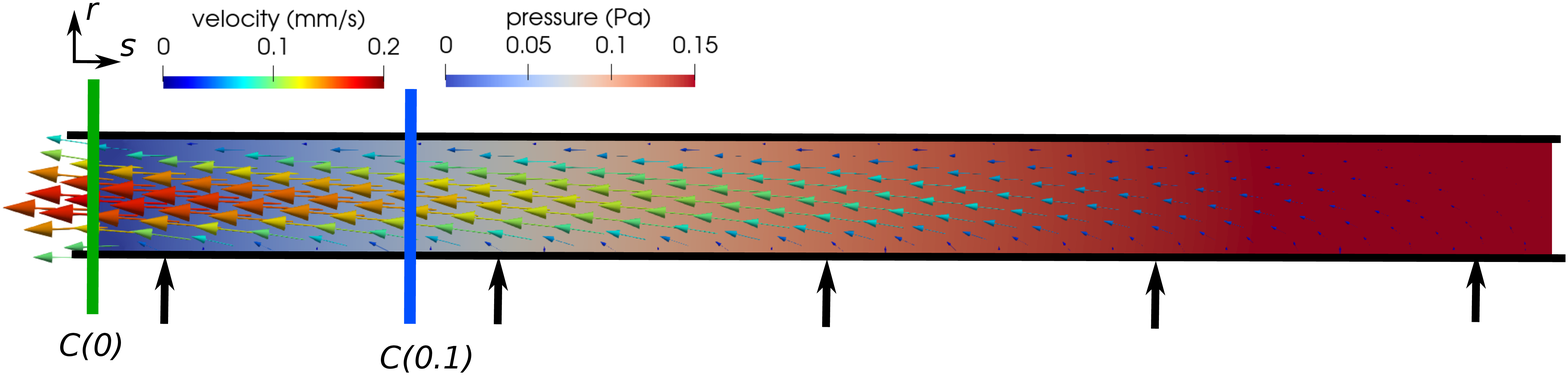}
     \subcaption{}  
     \label{fig:A2-vp}
   \end{subfigure}
   \begin{subfigure}{0.84\textwidth}
    \centering
    \includegraphics[width=0.89\textwidth]{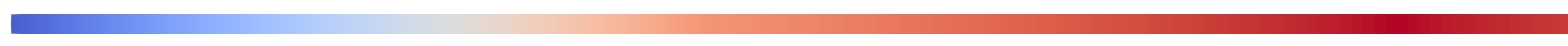}
    \includegraphics[width=0.9\textwidth]{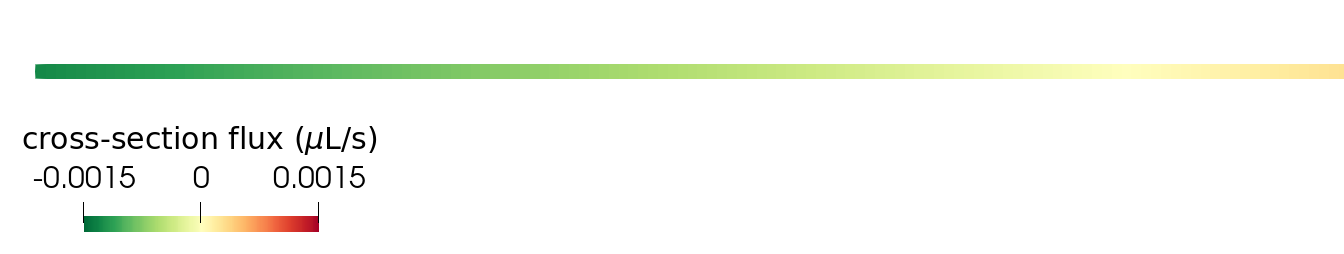}
     \subcaption{}  
     \label{fig:A2-qp}
   \end{subfigure}
    \iffalse
    \begin{subfigure}{0.3\textwidth}
    \includegraphics[width=0.9\textwidth]{Figures/A2-deltaR.png}
     \subcaption{}
     \label{fig:A2-deltaR}
     \end{subfigure}
     \fi
     \begin{subfigure}{0.49\textwidth}
        \includegraphics[width=0.8\textwidth]{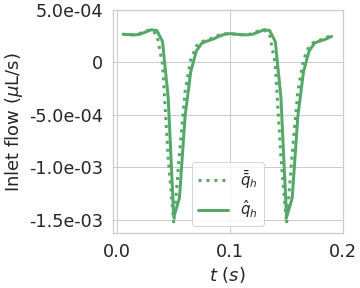}
        \subcaption{}
        \label{fig:A2-inlet}
     \end{subfigure}
     \begin{subfigure}{0.49\textwidth}
        \includegraphics[width=0.8\textwidth]{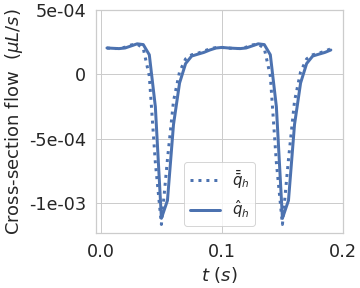}
        \subcaption{}
        \label{fig:A2-cs}
     \end{subfigure}
     \caption{PVS flux and pressure in an axisymmetric annular cylinder induced by a constant pressure difference or cardiac wall motion (Models A1, A2). (a) Model A1: A constant pressure gradient induces annular Poiseuille flow in both the full axisymmetric model (upper panel) and the reduced model (lower panel): snapshot of steady solution at $T=0.1$. (b-e) Model A2: Inner wall pulsations induce bidirectional and oscillatory flow. (b) Snapshot of the full model solutions at peak outflux ($t = 0.05$). Different cross-sections are marked in green (at the inlet) and blue (in the interior). (c) Pressure (upper panel) and cross-section flux $\bar{\bar{q}}_h$ (lower panel). (d) Cross-section flux predicted by the full model (dotted line) and the reduced model (solid line) at inlet versus time. (d) As for (c) but at the interior cross-section marked in (b).}

    \label{fig:results:A}
\end{figure}

\subsection{Reduced model exactly predicts pressure-driven axisymmetric flow characteristics}

Flow in an axisymmetric annular cylinder of length $\ell$ driven by a constant pressure difference $\Delta p$ (Model A1) is described by the analytic expression: 
\begin{align*}
    \hat{q}(s, t)&=A \frac{\Delta p}{\mu \alpha \ell} \left( 1-\exp \left(-\frac{\mu \alpha t}{\rho}\right)\right), \\ 
    \hat{p}(s, t)&=\frac{\Delta p}{\ell}s+\hat{p}(0),
\end{align*}
where $\alpha$ is the lumped flow parameter given by~\eqref{eq:alpha}  and which is constant in time and space in this case. For the velocity profile \eqref{eq:velo} defined over geometry A (cf.~Table \ref{tab:geometry-data}), $\alpha = 7325.3/\mathrm{m}^2$ , and $\mu \alpha / \rho = 5105.7 /\mathrm{s}$. Thus, the time-dependency is negligible after only a few milliseconds, and the flow develops near-instantaneously to steady-state Poiseuille flow.

Both the full and reduced models reproduce the exact annular Poiseuille flow characteristics of this case (Figure~\ref{fig:A1}). The numerical difference between the analytic and computed reduced solutions for the cross-section flux $\hat{q}$ and average pressure $\hat{p}$ is at machine precision ($\| \hat{q}(T) - \hat{q}_h(T)\| = 1.7 \times 10^{-14}$ and $\| \hat{p}(T) - \hat{p}_h(T)\| = 2.6 \times 10^{-17}$) ($T = 1$ s). In general, the total error is the sum of the model error and the numerical error associated with the space-time discrete approximation~\eqref{eq:variational}. For Model A1, the model error is zero as the model reduction assumptions (\textbf{I}--\textbf{V}) are exactly fulfilled by the geometry and flow pattern. As the total error also vanishes, we note that the numerical error is also negligible for this case.

\subsection{Reduced model accurately captures axisymmetric PVS wall pulsations}

Next, we examine the PVS flow and pressure generated by uniform axisymmetric pulsations of the inner PVS wall (Model A2, \Cref{fig:A2-vp}-\Cref{fig:A2-cs}). The inner wall movement changes the inner domain radius $R_1$ in time. The fluid is pushed out at the both ends as the PVS width decreases, and flows back in at both ends as the PVS width returns to baseline. This behaviour is reproduced by both the full (\Cref{fig:A2-vp}, \cite{daversin2020mechanisms}) and reduced models (\Cref{fig:A2-qp}). We note that the reduced model assumptions (\textbf{IV-V}) do not hold in this scenario as the PVS axial velocity profile is no longer identical to the Poiseuille velocity profile, and the pressure is not perfectly constant on each cross-section. Comparing the full and reduced cross-section fluxes $\bar{\bar{q}}_h$ and $\hat{q}_h$, we observe however that the two models still agree closely (\Cref{fig:A2-inlet}, \Cref{fig:A2-cs}), both at the inlet and at an interior cross-section. Moreover, the time-profile of the reduced and full cross-section flux approximations are very similar (both at the inlet and at the interior cross-section, \Cref{fig:A2-inlet}-\Cref{fig:A2-cs}), though with small ($\Delta t$ s) shifts in time. The peak outflux for the full model is $1.54 \times 10^{-3}$ $\mu$m$^3$/s, and $1.47 \times 10^{-3}$ $\mu$m$^3$/s for the reduced model, respectively (\Cref{fig:A2-inlet}). The peak pressure for the full model is $0.20$ Pa, and $0.19$ Pa for the reduced model. The relative model discrepancy in the peak cross-section flux (difference between the full and reduced peak flux) at the inlet is $4.1 \%$ and in the peak cross-section (average) pressure is $5.3 \%$. There is thus a small discrepancy between the two models, as expected by the violation of the reduced model assumptions. 

\subsection{Radial geometry variations induce small model errors}

\begin{figure}
    \centering
    \begin{subfigure}[b]{0.5\textwidth}
    \centering  
    \includegraphics[height=0.25\textheight]{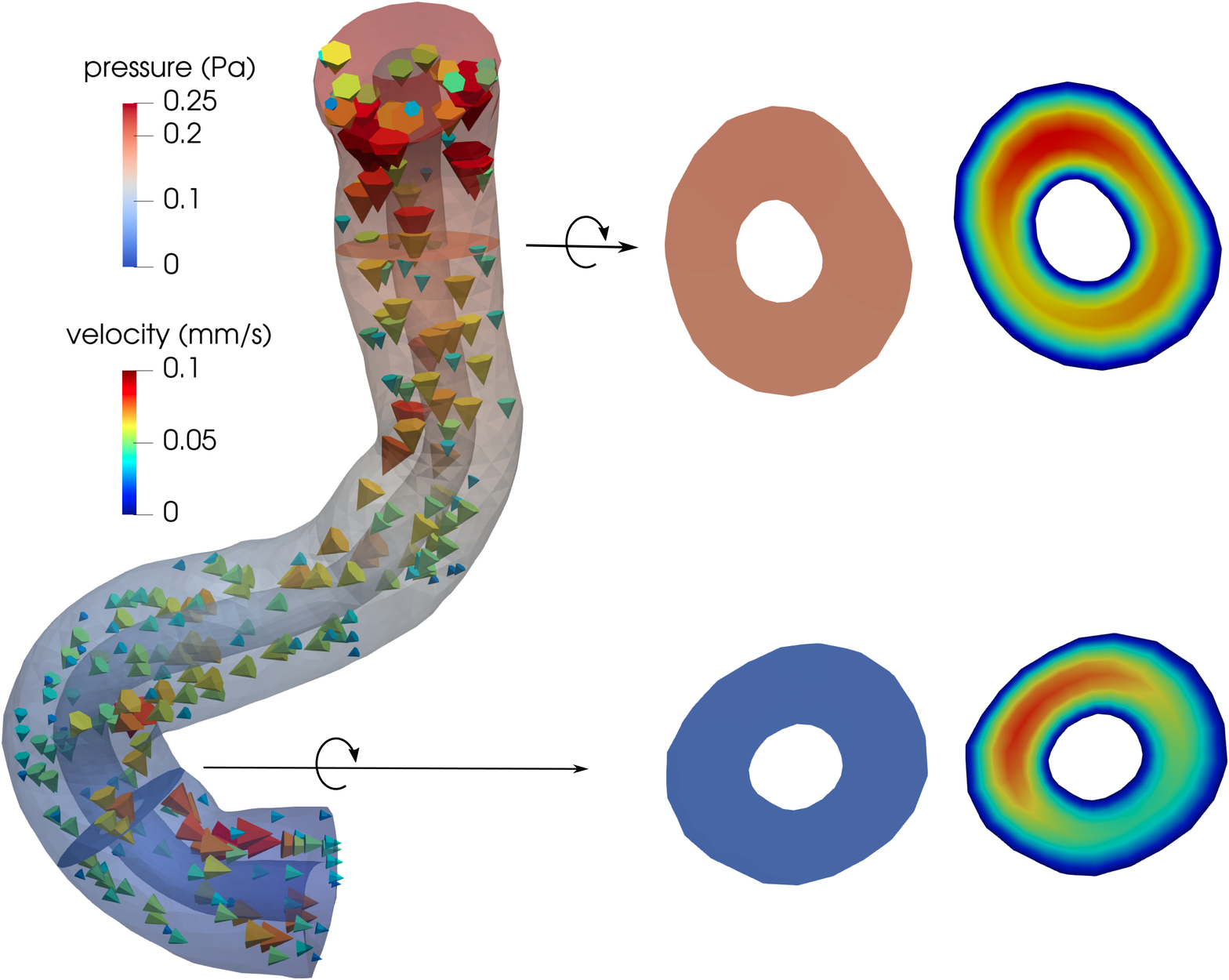}
    \caption{}
    \label{fig:modelB1-1}
    \end{subfigure}
    \begin{subfigure}[b]{0.45\textwidth}
      \centering
      \includegraphics[height=0.24\textheight]{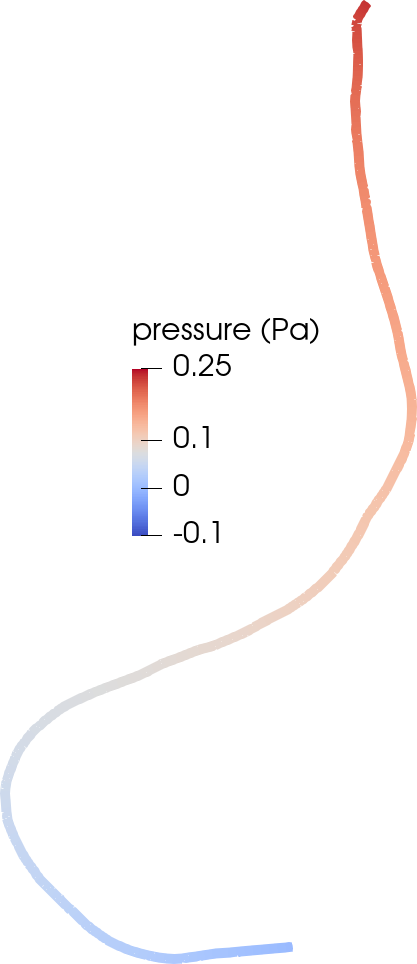}
      \includegraphics[height=0.24\textheight]{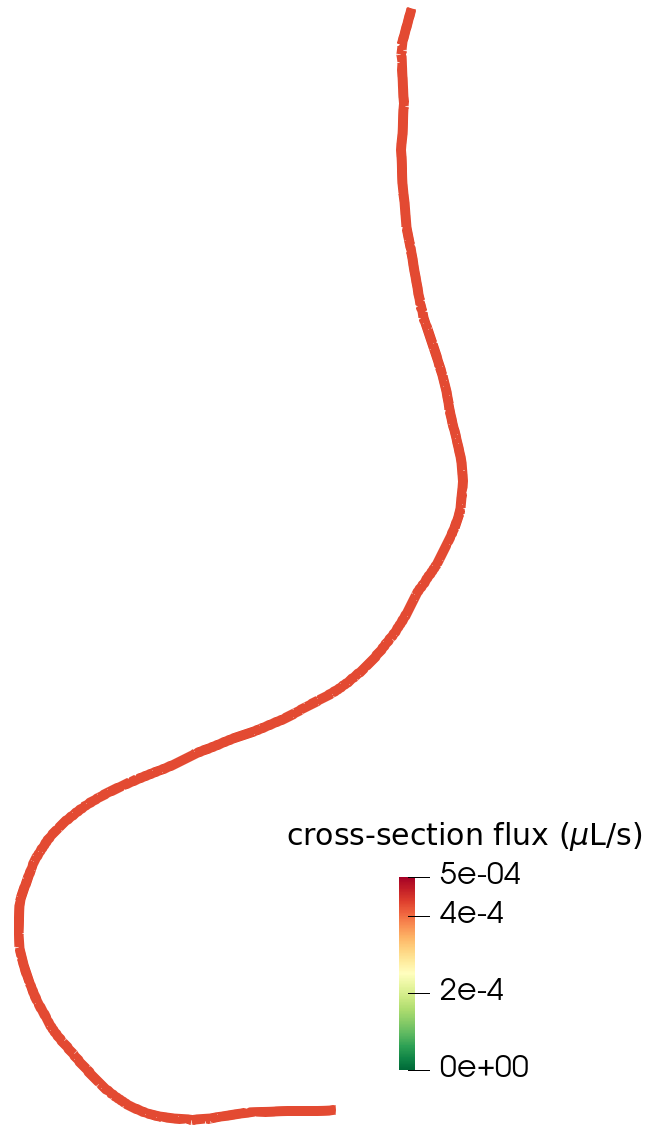}
      \caption{}
      \label{fig:modelB1-2}
    \end{subfigure}
    %\vspace{2em}
       \caption{In an image-based perivascular segment with varying radii, a pressure difference between inlet and outlet induces a pressure field that is nearly constant on each cross-section, but a velocity field that varies with the radial, angular and axial coordinates. (a) Full pressure and velocity approximations in the domain (left) along with close-up views of the pressure (middle) and velocity magnitude (right) at two cross-sections; (b) Reduced average cross-section pressure (left) and cross-section flux approximations (right).}
       \label{fig:modelB1}
    \end{figure}
In contrast to the axisymmetric geometry A, the image-based geometries B and C express angular and axial variations in radius. The inner and outer radii of these geometries vary along the length of the domain (with $s$) and depend on the angular coordinate $\theta$, with the latter violating model assumption \textbf{I}. To study the resulting model error in isolation, we again examine the pressure-driven flow predicted in full and reduced models but now of geometry B (Model B1, \Cref{fig:modelB1}). The full numerical approximation of the pressure is nearly constant over each cross-section . On the other hand, the velocity profile varies between cross-sections and with the angular coordinate within each cross-section (\Cref{fig:modelB1-1}). Therefore, we expect a larger model error in the reduced model compared to the previous case(s). At steady state ($t = 0.5$), the reduced pressure approximation $\hat{p}$ varies nearly linearly along the length of the domain as expected, and the reduced flux approximation $\hat{q}$ is essentially constant along the centerline with value $\hat{q} = 4.28 \times 10^{-4}$ $\mu$L/s. Computing the corresponding cross-section flux from the full model, we find values ranging from $3.5 \times 10^{-4}$ to $5.31 \times 10^{-4}$ $\mu$L/s. The total relative model discrepancy~\eqref{eq:modelerror} in the pressure $E_p = 2.6 \%$ and for the flux $E_q = 12.6\%$. 

\subsection{Reduced model is robust with respect to wall motion amplitude and frequency} 
\begin{figure}

%% Top row: 3D solution plot (left) and cross-sections (right)
    \begin{subfigure}[b]{0.5\textwidth}
      \centering
    \includegraphics[height=0.3\textheight]{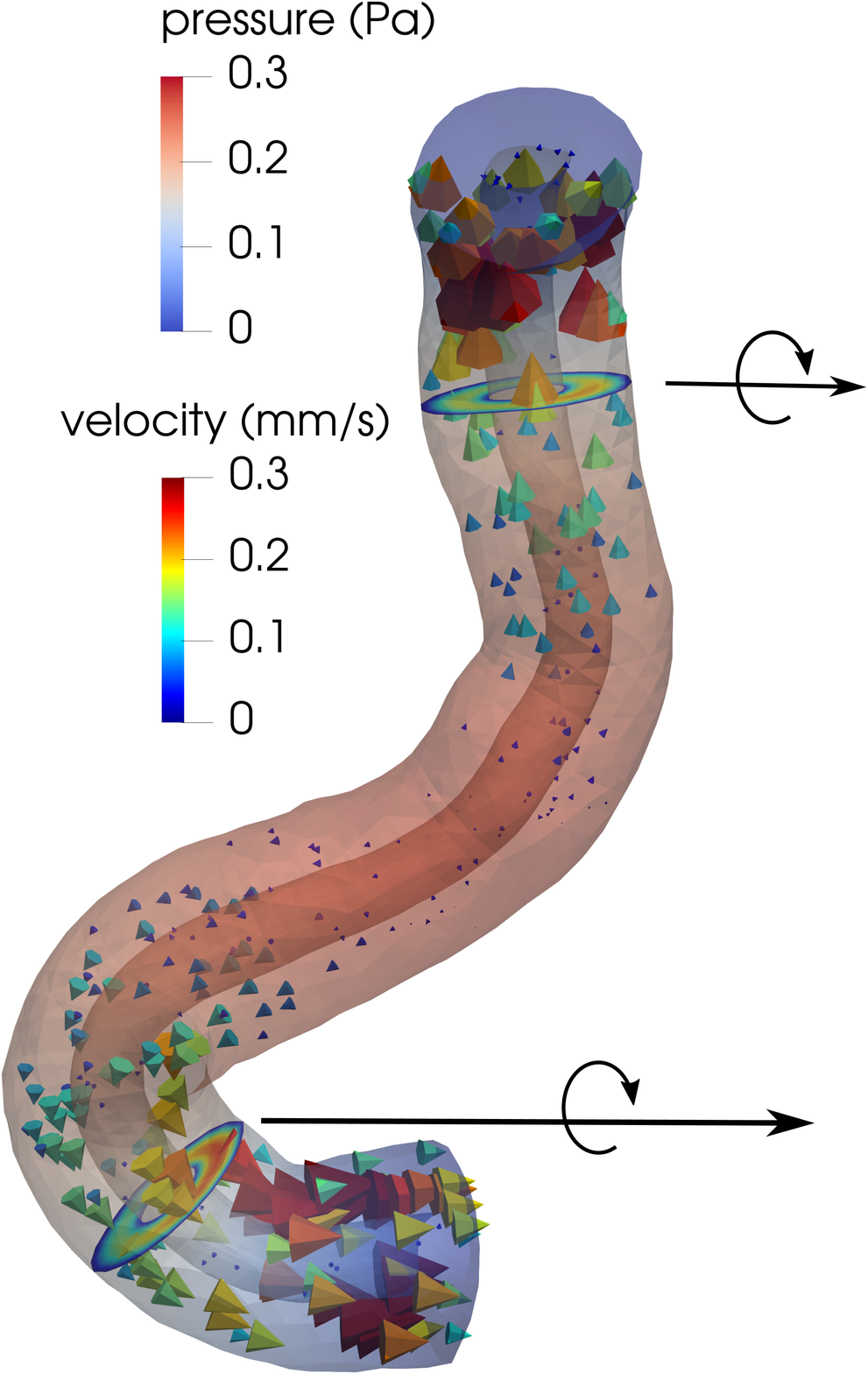}
    \hspace{-1em}
    \caption{}
     \label{fig:modelB2-1}
    \end{subfigure}
    \begin{subfigure}[b]{0.1\textwidth}
      \centering
    \vspace{2em}
    \includegraphics[width=0.99\textwidth]{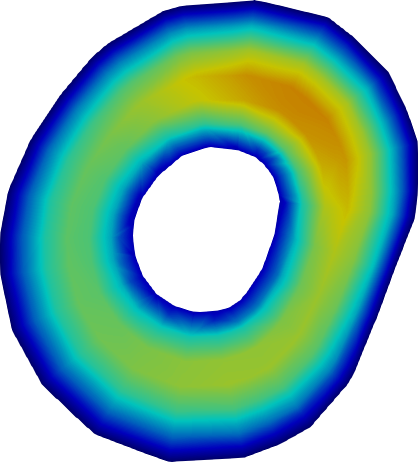}
    \vspace{6em}
    
    \includegraphics[width=0.99\textwidth]{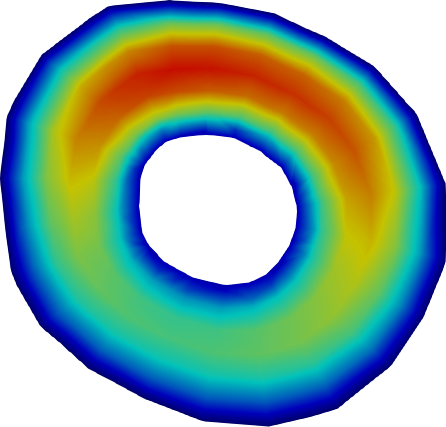}
    \vspace{1em}
    
    \caption{}
    \label{fig:modelB2-2}
    \end{subfigure}
    \begin{subfigure}[b]{0.1\textwidth}
      \centering
    \vspace{2em}
    \includegraphics[width=0.99\textwidth]{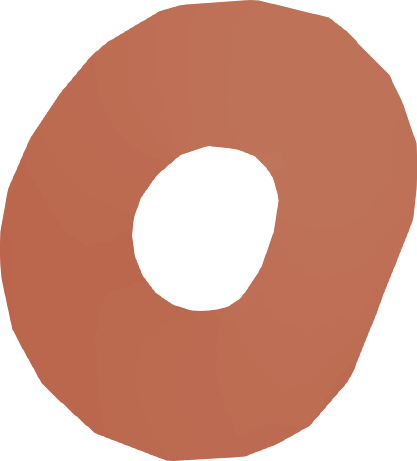}
    \vspace{6em}
    
    \includegraphics[width=0.99\textwidth]{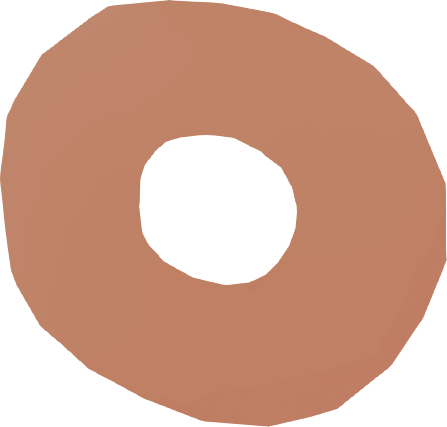}
    \vspace{1em}
    
    \caption{}
    \label{fig:modelB2-3}
    \end{subfigure}
    
    % Middle row: Comparison of 1D and 3D fluxes (left) and pressures (right)
    \begin{subfigure}[b]{0.45\textwidth}
    \centering
    \includegraphics[height=0.23\textheight]{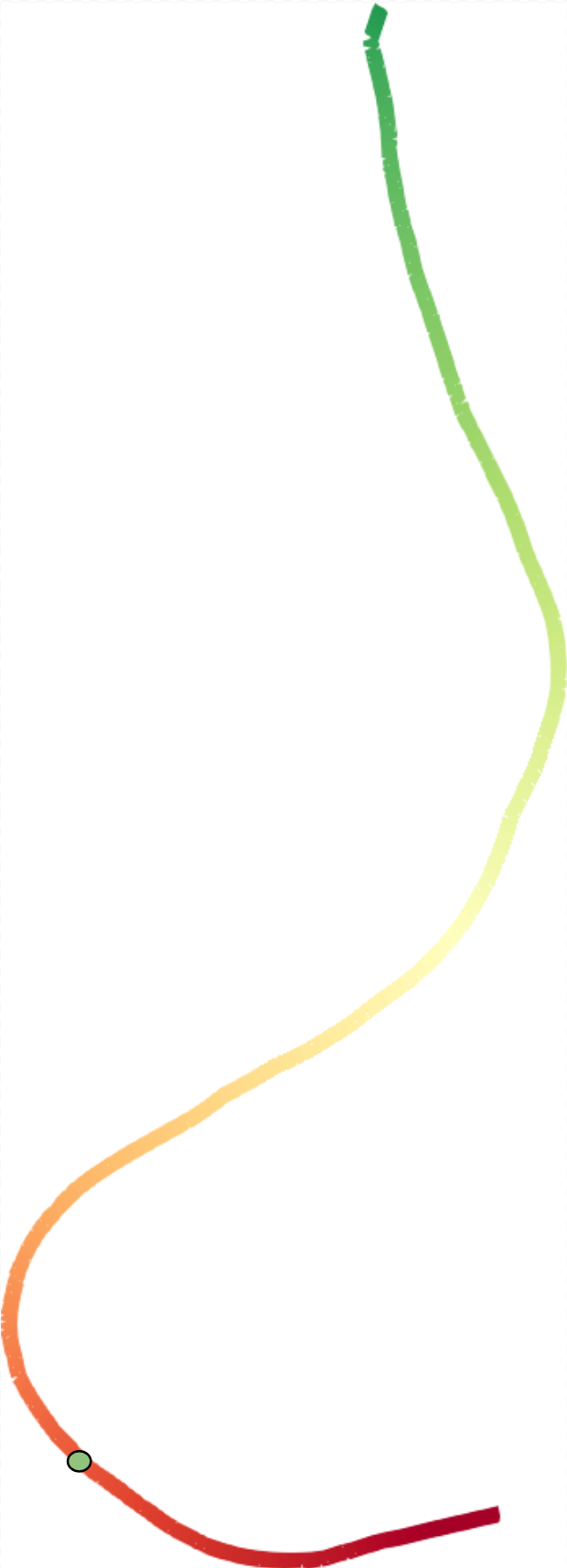}
    \includegraphics[height=0.23\textheight]{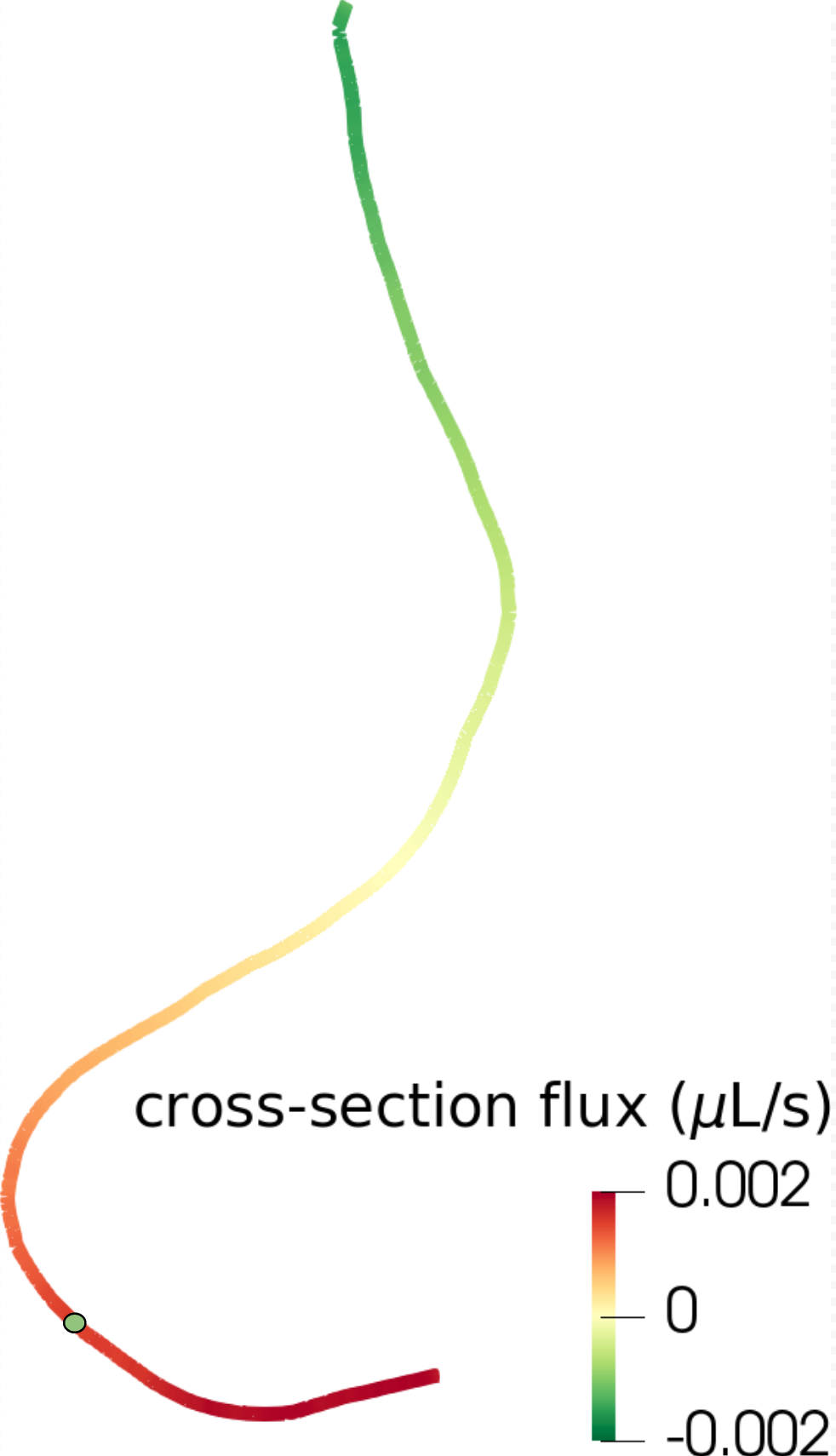}
    \caption{}
    \label{fig:modelB2-4}
    \end{subfigure}
    \begin{subfigure}[b]{0.45\textwidth}
    \centering
    \includegraphics[height=0.23\textheight]{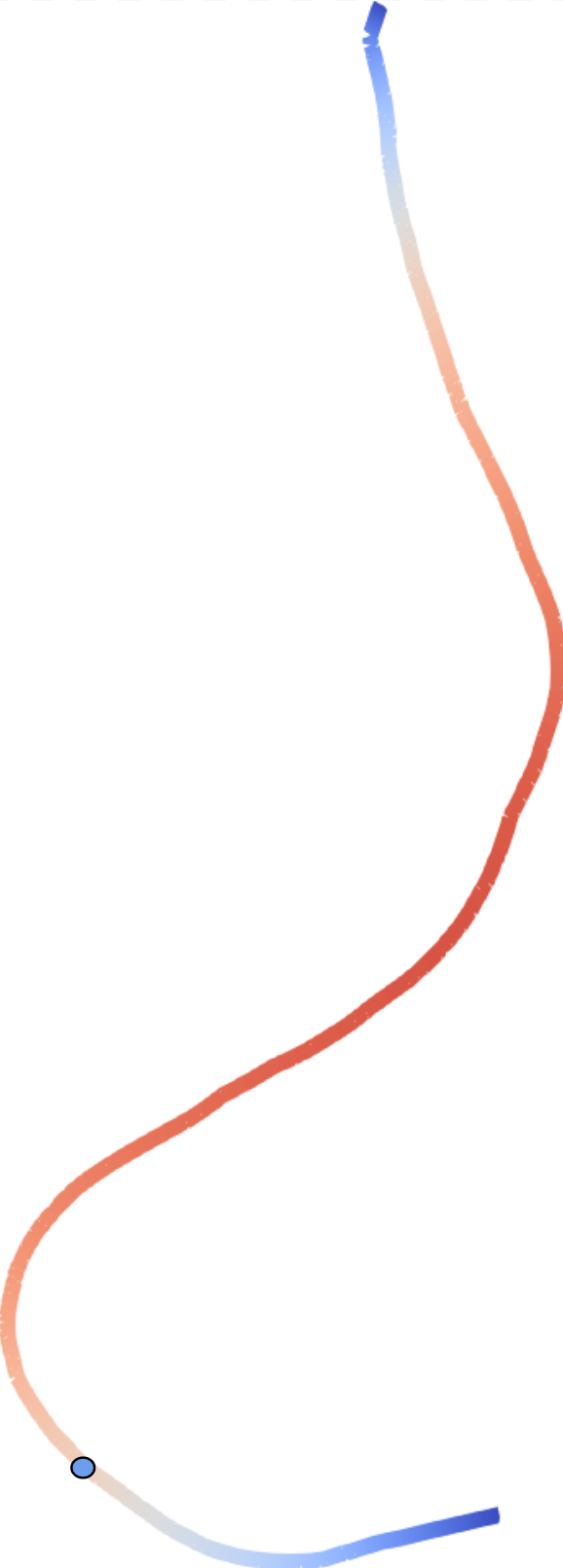}
    \includegraphics[height=0.23\textheight]{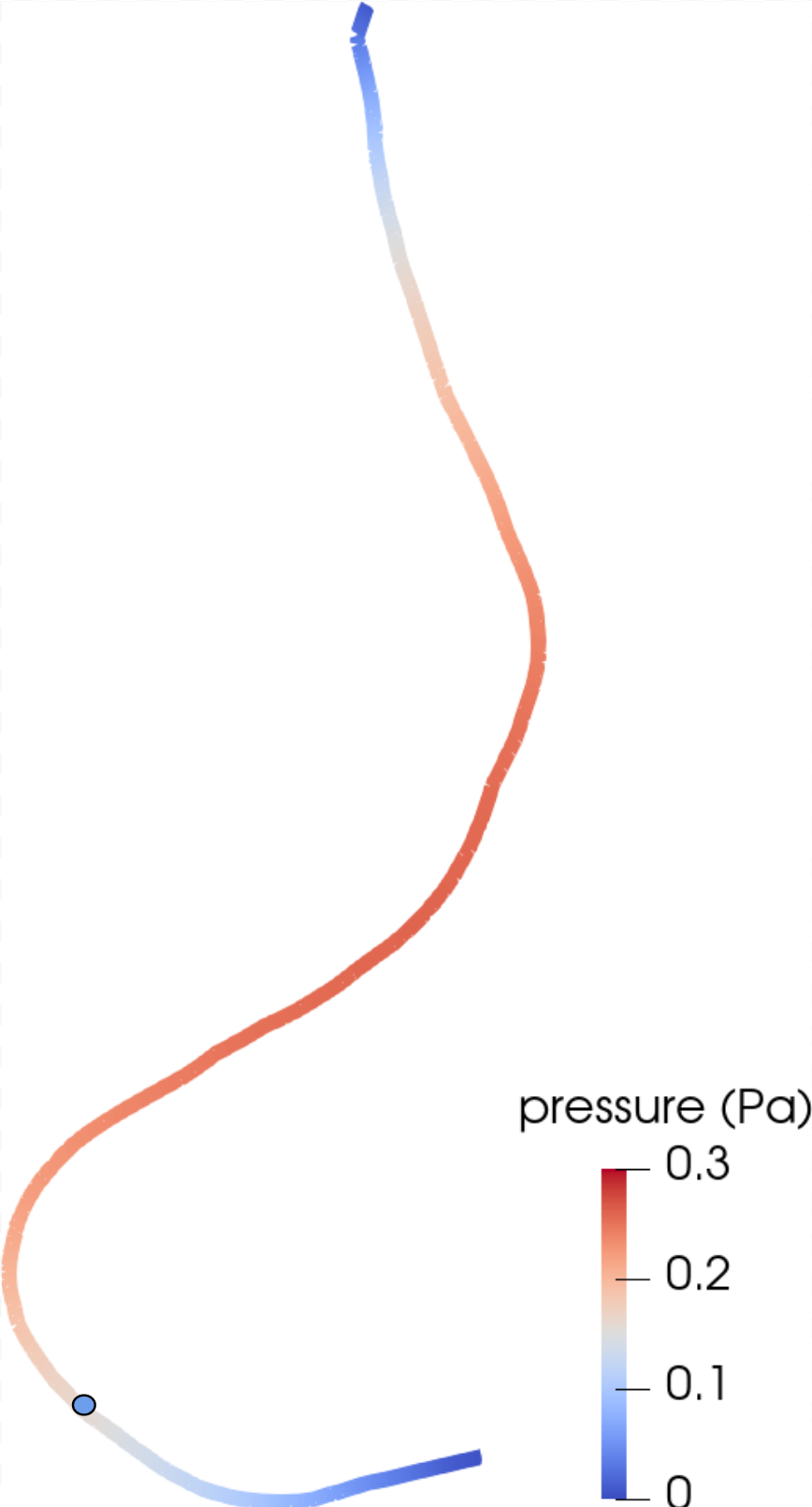}
    \caption{}
    \label{fig:modelB2-5}
    \end{subfigure}
    \begin{subfigure}{0.45\textwidth}
    \centering
    \includegraphics[height=5cm]{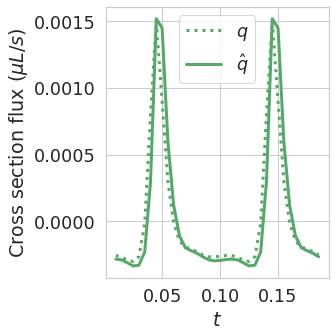}
    \caption{}
    \label{fig:modelB2-6}
    \end{subfigure}
    \begin{subfigure}{0.45\textwidth}
    \centering
    \includegraphics[height=5cm]{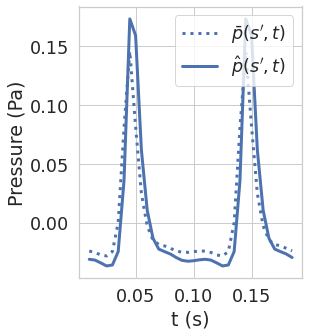}
    \caption{}
    \label{fig:modelB2-7}
    \end{subfigure}

    \caption{Cardiac wall motion induce substantial pulsatile pressures and velocities in an image-based perivascular space segment, with the reduced model accurately capturing flow, pressure and transport characteristics. (a) Snapshot of the pressure and velocity at time of peak pressure ($t = 0.05$s); (b) Velocity at upper and lower cross-section (zoom of (a)); (c) Pressure at upper and lower cross-sections (zoom of (a)); (d) Cross-section flux from reduced model (left) and full model (right); (e) cross-section average pressure from reduced model (left) and full model (right); (f) full and reduced model cross-section fluxes at the lower cross-section over time; (g) full and reduced model pressures at the lower cross-section over time.}
    \label{fig:modelB}
\end{figure}

Cardiac wall motion and vasomotion may drive pulsatile perivascular flow with different flow characteristics. To evaluate the model discrepancy induced by different physiological drivers, we compare the full and reduced models over an image-based PVS segment driven by wall motion induced by the cardiac pulse wave (Model B2) and by vasomotion (Model B3). The cardiac pulse wave induces wall motion at a higher frequency (10 Hz) travelling at a higher wave speed ($1000$ mm/s), while vasomotion creates pulsations at lower frequencies ($0.1$ Hz) and at a lower wave speed ($0.8$ mm/s). Both models include angularly, axially and temporally varying radii, and we expect model assumptions \textbf{I}, \textbf{IV-V} to not hold.  

\begin{figure}
    \centering
      \begin{subfigure}[b]{0.99\textwidth}                
      \includegraphics[width=0.9\textwidth]{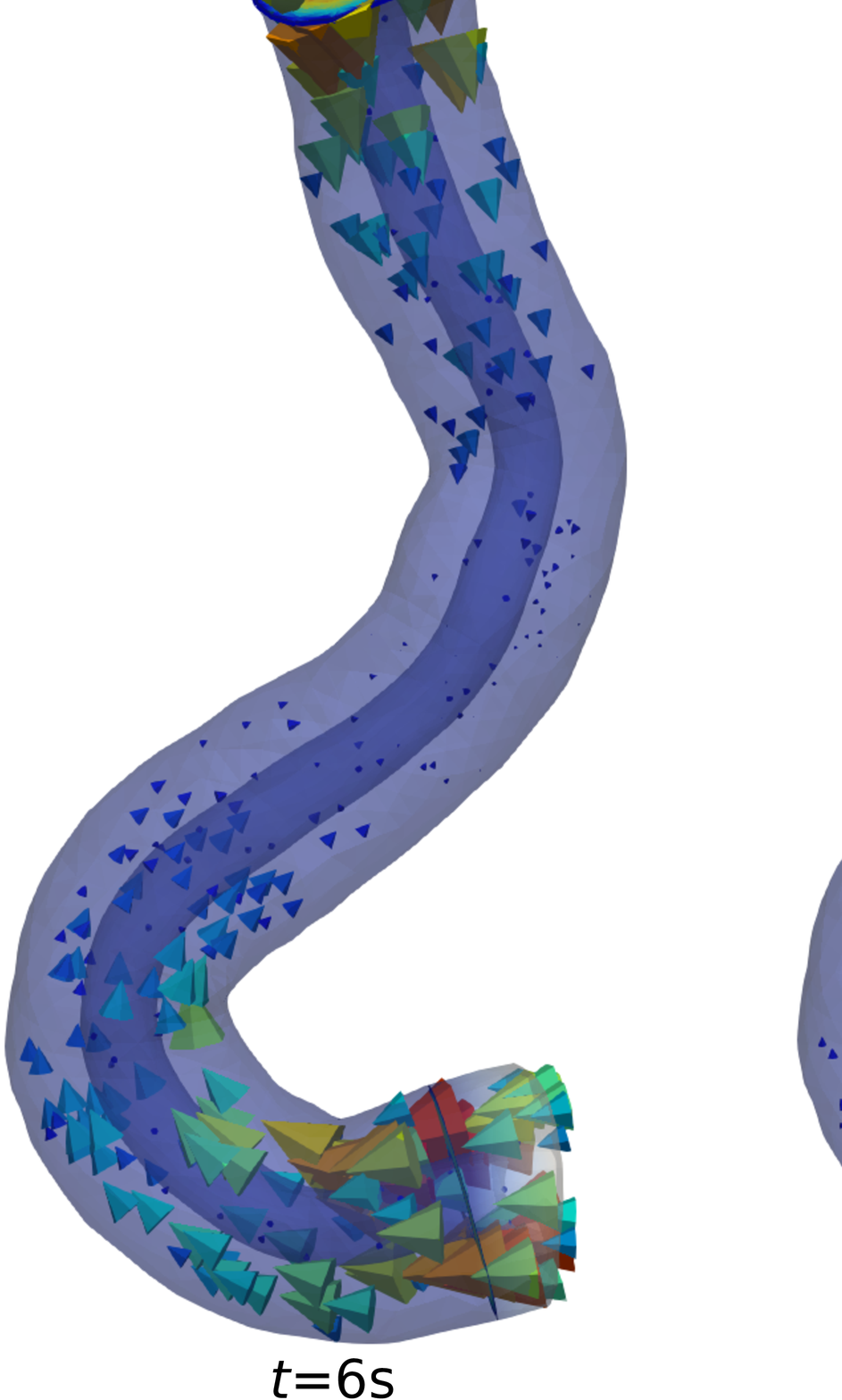}
    \caption{}
    \label{fig:modelB3-1}
    \end{subfigure}

    \begin{subfigure}[b]{0.99\textwidth}
    \includegraphics[width=0.99\textwidth]{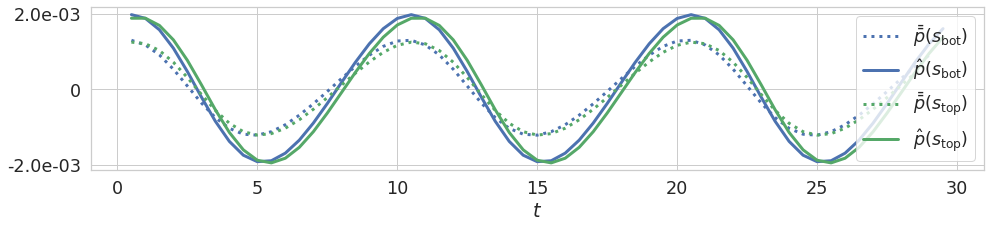}
    \includegraphics[width=0.99\textwidth]{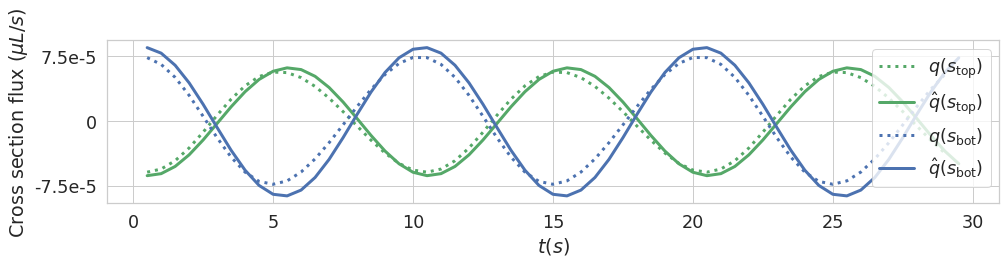}
    \caption{}
    \label{fig:modelB3-2}
    \end{subfigure}
    \caption{Vasomotion induces higher domain deformations but lower wall velocities, pressure differences and cross-section fluxes. (a) Snapshots of the full model pressure and velocity at different time points with cross-section velocities (top). (b) Average pressure (upper panel) and flux (lower panel) for the full and reduced models at upper and lower cross-sections over time. The values at the different cross-sections are slightly shifted in time due to the travelling vasomotion. The pressure model discrepancy dominates the flux differences.}
    \label{fig:modelB3}
\end{figure}

Both pairs of models induce pulsatile bidirectional flow in and out of the PVS segment in synchrony with the pulsating wall (\Cref{fig:modelB}, Supplementary Video S1) with peak pressure magnitude in the middle of the segment, and conversely, low velocities in the middle of the domain and higher velocities near the PVS ends. Both model scenarios lead to pressure fields that are nearly constant on each cross-section (\Cref{fig:modelB2-3}, \Cref{fig:modelB3}), but with angularly varying velocity profiles (\Cref{fig:modelB2-2}, \Cref{fig:modelB3}). 

For the cardiac wall motion, the cross-section average of the full pressure $\bar{\bar{p}}_h$ ranges from $-0.05$ to $0.26$ Pa, while the full cross-section flux $\bar{\bar{v}}_h$ ranges from $-1.54 \times 10^{-3}$ to  $1.95 \times 10^{-3}$ $\mu$L/s. The reduced model accurately captures the temporal and spatial characteristics of the full model (\Cref{fig:modelB2-4}--\Cref{fig:modelB2-7}). For the reduced model, the cross-section pressure $\hat{p}_h$ ranges from $-0.06$ to $0.29$ Pa, while the cross-section flux $\hat{q}_h$ is between $-1.61 \times 10^{-3}$ and $2.23 \times 10^{-3}$ $\mu$L/s. Comparing the full and reduced pressure and flux over time at an interior, lower cross-section, we observe that the reduced model slightly overestimates the peak pressure and flux when compared to the full model (\Cref{fig:modelB2-5}, \Cref{fig:modelB2-6}). The relative difference in peak positive pressure between the two models at this cross-section is $19 \%$ and $30\%$ in peak negative pressure. For the flux, the corresponding model discrepancies are $1.2\%$ and $11\%$.

For the vasomotion scenario, the domain movement is larger compared to the cardiac wall motion, but the wall velocity is lower (peak wall speed of $0.001$m/s vs $0.005$ mm/s). The resulting peak (in terms of magnitude) cross-section pressure is $-0.012$ Pa and peak cross-section flux is $9.14 \times 10^{-5}$ $\mu$L/s (\Cref{fig:modelB3}). These are one-to-two orders of magnitude lower than for the cardiac wall motion scenario. Comparing the full and reduced models in two interior (upper and lower) cross-sections, we observe that the cross-section pressure $\hat{q}_h$ matches pulsatile behaviour of the average cross-section pressure in the full model $\bar{\bar{q}}_h$ (\Cref{fig:modelB3-2}) but that the peak amplitude is higher. The largest model differences in pressure at lower cross-section is at the peak pressure; there the relative difference in peak pressure is $52\%$. The similar observations hold for the flux, but the model discrepancies are lower: the relative difference in peak flux is $15 \%$. Moreover, the full and reduced models agree on a pressure phase shift of $0.5$s. In agreement with our previous findings, the reduced pressure approximation displays a greater model discrepancy with higher predicted pressure variations in the reduced model (\Cref{fig:modelB3-2}).

\subsection{Reduced model captures flow and transport characteristics through bifurcations}

\begin{figure}
    \centering
    \begin{subfigure}[b]{0.54\textwidth}
        \includegraphics[height=0.95\textwidth]{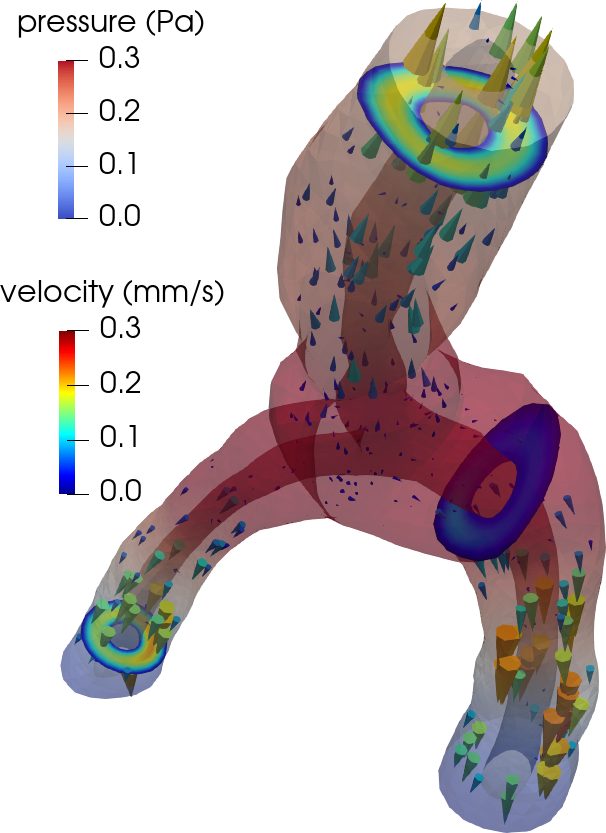}
        \caption{}
        \label{fig:modelC12-a}
    \end{subfigure}
    \hspace{-1.5em}
    \begin{subfigure}[b]{0.44\textwidth}
        \includegraphics[width=0.9\textwidth]{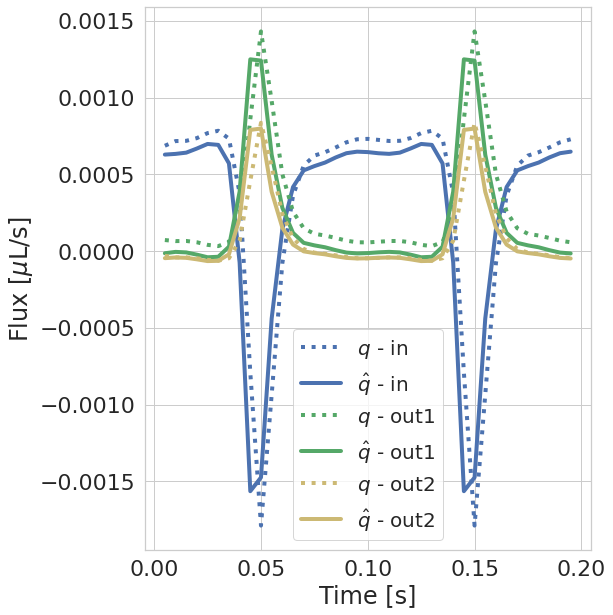}
        \vspace*{-0.25em}
        \caption{}
        \label{fig:modelC12-b}
    \end{subfigure}
    % Pressure
    \begin{subfigure}[b]{0.22\textwidth}
        \includegraphics[height=0.2\textheight]{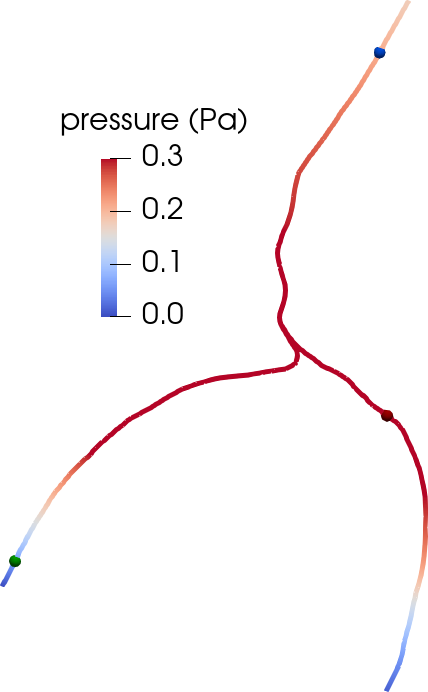}
        \caption{}
        \label{fig:modelC12-c}
    \end{subfigure}
    \begin{subfigure}[b]{0.22\textwidth}
        \includegraphics[height=0.2\textheight]{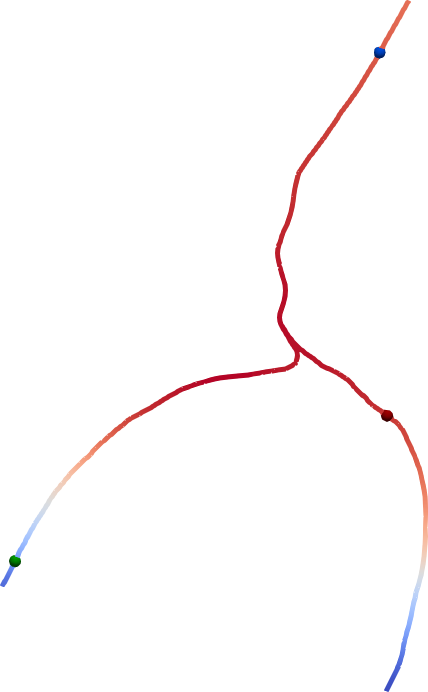}
        \caption{}
        \label{fig:modelC12-d}
    \end{subfigure}
    \begin{subfigure}[b]{0.1\textwidth}
        \includegraphics[width=0.97\textwidth]{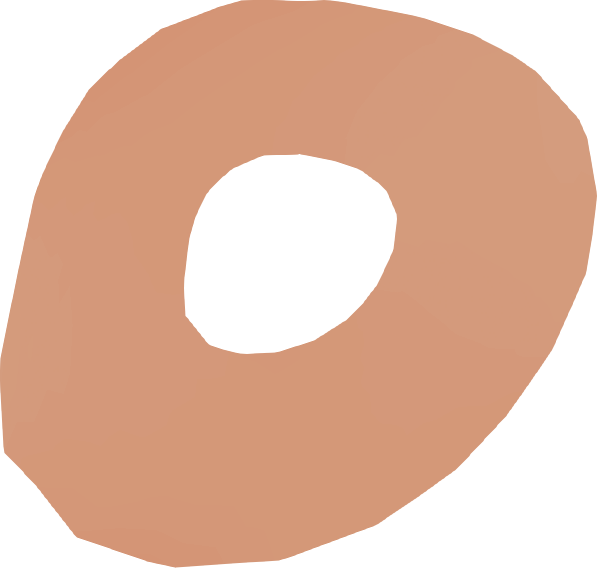}
        \includegraphics[width=0.97\textwidth]{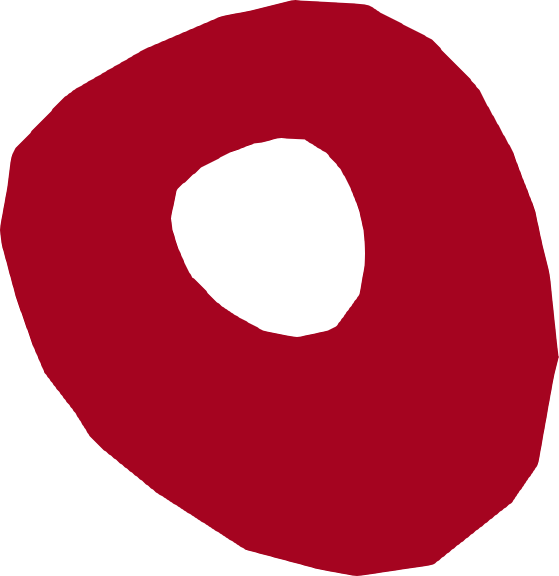}
        \includegraphics[width=0.97\textwidth]{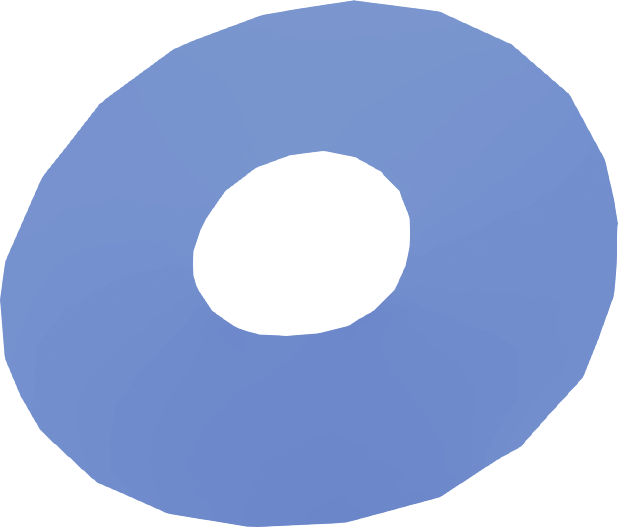}
        \caption{}
        \label{fig:modelC12-e}
    \end{subfigure}
    \begin{subfigure}[b]{0.44\textwidth}
        \hspace{0.8em}
        \includegraphics[width=0.8\textwidth]{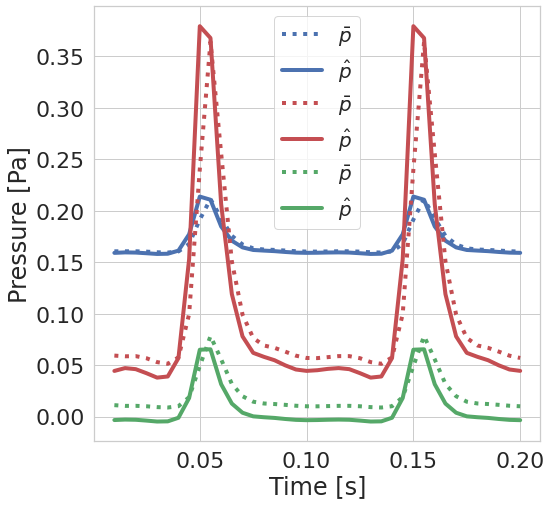}
        \vspace*{-0.5em}
        \caption{}
        \label{fig:modelC12-f}
    \end{subfigure}
    % Flux
    \begin{subfigure}[b]{0.22\textwidth}
        \includegraphics[height=0.2\textheight]{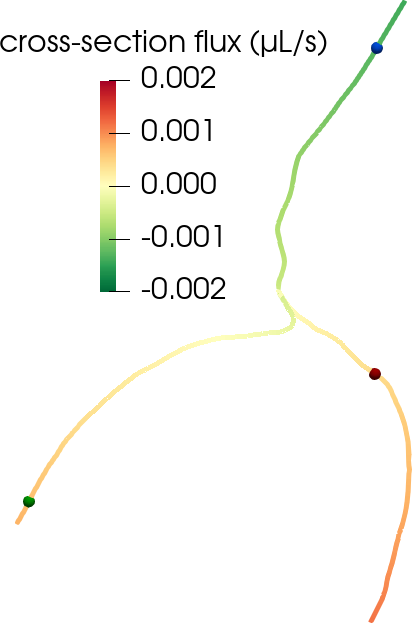}
        \caption{}
        \label{fig:modelC12-g}
    \end{subfigure}
    \begin{subfigure}[b]{0.22\textwidth}
        \includegraphics[height=0.2\textheight]{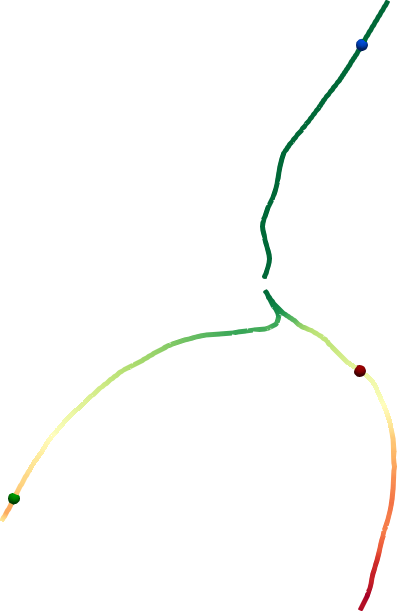}
        \caption{}
        \label{fig:modelC12-h}
    \end{subfigure}
    \begin{subfigure}[b]{0.1\textwidth}
        \includegraphics[width=0.97\textwidth]{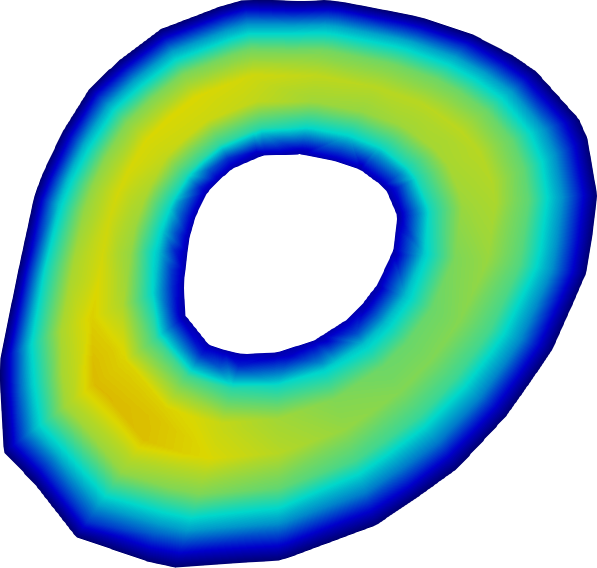}
        \includegraphics[width=0.97\textwidth]{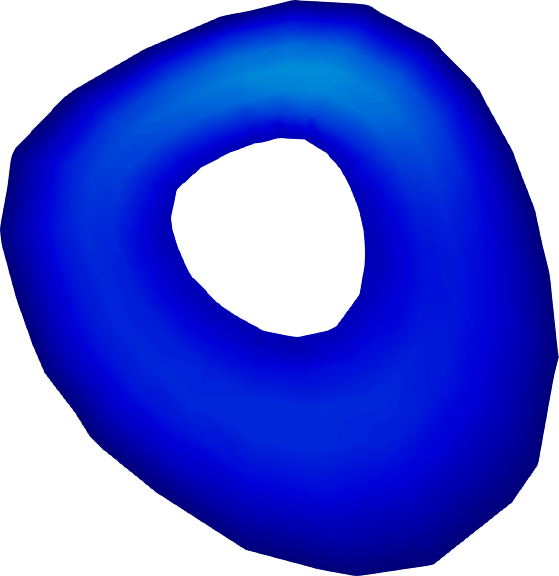}
        \includegraphics[width=0.97\textwidth]{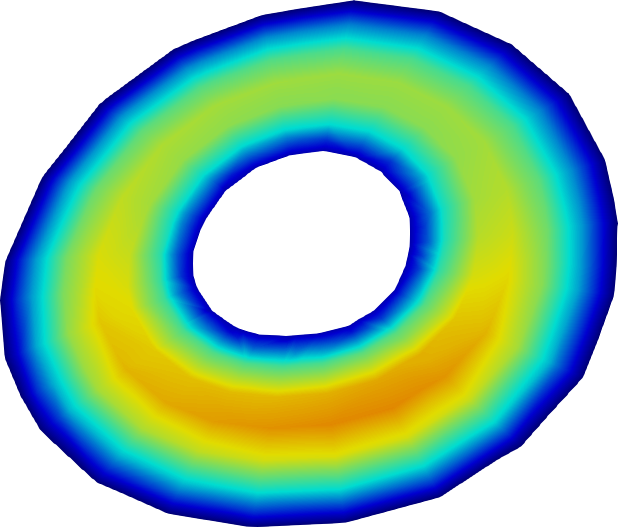}
        \caption{}
        \label{fig:modelC12-i}
    \end{subfigure}
    \begin{subfigure}[b]{0.44\textwidth}
        \includegraphics[width=0.85\textwidth]{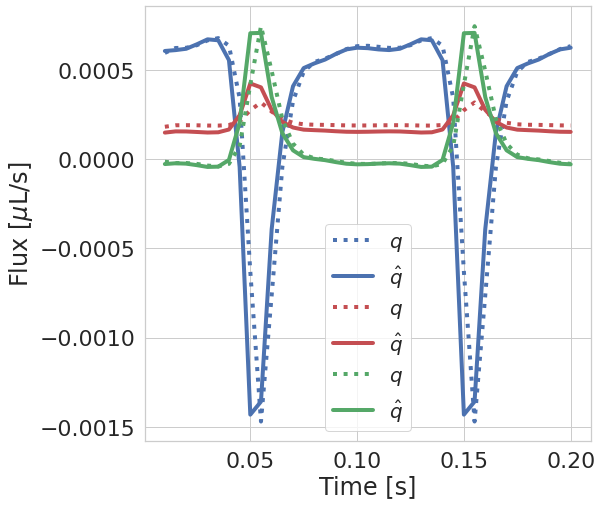}
        \vspace*{-0.5em}
        \caption{}
        \label{fig:modelC12-j}
    \end{subfigure}
    \caption{Flow through a bifurcating PVS (Model C12) (a) Snapshot of pressure and velocity from full model at peak velocity ($t = 0.05$). (b) Full versus reduced cross-section flux at inlet (in) and outlets (out1 and out2) over time. (c) Snapshot of reduced cross-section pressure at peak velocity. (d) Snapshot of average cross-section pressure at the same time. (e) Pressure at upper, middle and lower cross-sections (zoom of (a)). (f) Full versus reduced cross-section average pressure at cross-sections. (g) Snapshot of reduced cross-section flux at peak velocity ($t = 0.05$). (h) Snapshot of cross-section flux from the full model at the same time. (i) Flux at upper, middle and lower cross-sections (zoom of (a)). (j) Full versus reduced cross-section flux at cross-sections.}
    \label{fig:modelC12}
\end{figure}

Now, we turn to compare the full and reduced model predictions of physiologically realistic perivascular flow in an image-based PVS surrounding a vascular bifurcation (Model C12). The prescribed pressure difference between inlet and outlets as well as the cardiac wall motion induces pulsatile flow with a net flow component~\cite{daversin2020mechanisms} (\Cref{fig:modelC12-a}, Supplementary video S2). We note that the domain radii vary both angularly and axially, also for the initial domain, and also that the presence of a bifurcation region induces non-Poiseuille/non-Womersley-type velocity profiles. Comparing the full and reduced average pressure and flux at the time of peak velocity (\Cref{fig:modelC12-f}, \Cref{fig:modelC12-j}), we note that the reduced model captures the qualitative and quantitative flow and pressure characteristics. The bifurcation conditions are satisfied at the bifurcation point $b$ (\Cref{fig:modelC12-c}, \Cref{fig:modelC12-g}) with a parent branch flux $\hat{q}(b)|_{\Lambda^P} = -5.52 \times 10^{-4} \mu$L/s and daughter branch fluxes $\hat{q}(b)|_{\Lambda^{d_1}} = -7.8 \times 10^{-5} \mu$L/s and $\hat{q}(b)|_{\Lambda^{d_2}} = -4.74 \times 10^{-4} \mu$L/s. The uneven flux distribution is induced by the smaller average width of one of the daughter vessels. The predicted stress $\hat{\sigma}$ is continuous (data not shown). 

The reduced peak cross-section flux (over time) at the inlet is $-1.5 \times 10^{-3} \mu$L/s, and $1.2 \times 10^{-3} \mu$L/s and $7.9 \times 10^{-4} \mu$L/s at the outlets (\Cref{fig:modelC12-b}). Comparing the peak flux model discrepancies at the inlet and outlets, we note that the discrepancy is largest at larger daughter outlet with a relative difference of $12 \%$. Comparing the full and reduced peak pressures at the upper, middle and lower cross-sections, we find relative model differences of $1 \%$, $4 \%$, and $16 \%$. The analogous numbers for the fluxes are $1 \%$, $34 \%$, and $5 \%$. Thus, the model discrepancies for the flux are larger near the bifurcation region (\Cref{fig:modelC12-f}, \Cref{fig:modelC12-j}). 

The net flow is a key quantity of interest for the physiological relevance of perivascular flow and transport. The net flow per cycle in the full model is $3.5 \times 10^{-5} \mu$L, and $2.9 \times 10^{-5} \mu$L for the reduced model, corresponding to a relative difference of $17\%$.

\subsection{Reduced models offer orders of magnitude saving in computational resources}

\begin{table}
\centering
\begin{tabular}{c|rr|rr|rr}
\toprule    
     & \multicolumn{2}{c}{d.o.fs} & \multicolumn{2}{c}{time (s)}  & \multicolumn{2}{c}{memory (MB)}  \\    
    Model & Full & Reduced & Full & Reduced & Full & Reduced \\
    \midrule
    A2 & 9 103 & 194 & 0.16 & 0.35 & 180  & 146 \\
    B2 & 287 432 & 1067 & 42.33 & 0.83 & 6261 & 133 \\
    C12 & 401 156 & 749 & 130.57 & 0.76 & 8874 & 176 \\
    \bottomrule
\end{tabular}
\caption{The geometrically-reduced models reduce computational cost by orders of magnitude. Number of degrees of freedom $d.o.f.s$, computational time (average time for a single time step) and memory usage (peak memory usage throughout the simulation) for the full models (2D/3D) and reduced models (1D).} 
\label{tab:computational}
\end{table}
Accurate direct three-dimensional simulations of pulsatile perivascular fluid flow in large, deforming vascular networks involve a significant computational cost. The expense is dominated by solving large linear systems of equations at each time step. For instance, even the moderate-resolution single-bifurcation model considered here (model C12) includes more than 17 000 vertices, 88 000 mesh cells and 400 000 degrees of freedom. For a small-scale idealized model such as axisymmetric Model A2, the reduced model uses $2.1\%$ of the number of degrees of freedom but approximately the same amount of memory and longer runtime (0.16 vs 0.35 s per time step, \Cref{tab:computational}). However, the one-dimensional models reduce computational cost substantially for the image-based geometries (\Cref{tab:computational}). For the image-based perivascular segment (Model B2), the reduced model uses $0.4\%$ of the number of degrees of freedom, $2.0\%$ of the runtime, and $2.1\%$ of the memory of the full model. For the image-based bifurcating PVS (Model C12), the reduced model uses $0.18\%$ of the number of degrees of freedom, $0.6\%$ of the runtime and $2.0\%$ of the memory of the full model. Overall, the reduced model reduces the computational expense, both in terms of computational time and memory, by several orders of magnitude for image-based PVS segments.

\section*{Discussion}

We have proposed a new mathematical and numerical framework based on topological and geometrical model reduction for computational modelling and simulation of steady and pulsatile fluid flow in deformable perivascular space networks. The reduced model is defined over a perivascular centerline network and predicts the fluid flux and average pressure in each cross-section of each network branch. By numerically comparing direct three-dimensional simulations of the fluid flow with the reduced model results for a range of physiological scenarios, we find that the reduced model accurately captures the important flow characteristics with cross-section peak pressure discrepancies ranging from $0 \%$ to $52 \%$ and peak flux discrepancies ranging from $0 \%$ to $35 \%$. Our findings indicate that reduced model is robust with respect to physiologically relevant spatial and temporal variations in the vascular radius. Moreover and importantly, the computational cost of the reduced model is several orders of magnitude lower than that of the corresponding full model. 

While geometrically-reduced network models of pulsatile blood flow have become a standard computational tool~\cite{olufsen1999structured, olufsen2004, coccarelli2021framework}, network models of perivascular fluid flow have mainly focused either on quantifying flow resistance~\cite{faghih2018bulk, tithof2019hydraulic} or predicting steady flow~\cite{tithof2021network}. In the latter, Tithof et al present the results of a network model of glymphatic flow under different parameters, using resistance models to compute flow in idealized domains. For the open channel flow, they compute the flow therein via Darcy's law $v = -(\kappa A / \nu) \nabla p$ with permeability 
\begin{align}
    \kappa = \frac{1}{8}\left( R_2^2+R_1^2-\frac{R_2^2-R_1^2}{\ln(R_2/R_1)}\right) .
    \label{eq:kappa:tithof}
\end{align}
This relationship holds under the assumption of Poiseuille flow in the open, annular channel (for which there is an analytic solution) and corresponds to the permeability required for this solution to satisfy Darcy's law. For steady-state flow ($\partial_t v = \partial_{ss} v = 0$) driven by a constant pressure difference, the reduced model equations~\eqref{eq:1d-pvs} simplify to the Darcy flow equation with permeability
\begin{align}
    \kappa = \frac{1}{\alpha} .
    \label{eq:kappa}
\end{align}
In the idealized Model A1 scenario, the two definitions of $\kappa$ (\eqref{eq:kappa:tithof} and~\eqref{eq:kappa}) agree, with $\kappa = 1.36 \times 10^{-4}$ mm$^2$, and thus the models coincide within this regime. 

Rey and Sarntinoranont \cite{rey2018pulsatile} also introduced two hydraulic models to predict fluid flow induced by blood pressure wave pulsations, and in particular net flow and transport. Their models also capture the pulsatile flow generated by the volume changes induced by a pulsating inner boundary, but under other modelling assumptions and without considering bifurcations, and thus differ from the one considered here. However, their peak fluid velocities of the order tens of $\mu$m/s is of the same order as the fluid velocities predicted in single branches here (Models A2, B2, B3), as are the pressures on the order of up to $0.3$ Pa.

Several different bifurcation conditions have been proposed in the literature. In one-dimensional blood flow models, the most common conditions are conservation of flux combined with continuity of pressure~\cite{olufsen2004, notaro2016mixed}. These conditions may be imposed directly on the pressure and flux solution variables~\cite{olufsen2004}, or weakly in the variational formulation~\cite{notaro2016mixed}. Here, we also enforce conservation of flux, but in place of the strong pressure continuity condition, we weakly impose the continuity of the normal stress. This approach gives a natural setting for Stokes flow and allows for a compatible variational formulation using a Lagrange multiplier space.  

In terms of limitations, we here focus on models of perivascular flow and the effect of vascular pulsations on perivascular flow, and not on the full interplay between vascular, perivascular and interstitial flow and deformation, nor on the transfer across the blood-brain barrier or the glial limitans. For healthy arterial and venous regions, in which the blood flow dynamics dominate the perivascular flow and pressure, we expect this one-way (vascular-to-perivascular) coupling to capture the leading order dynamics. Moreover, in light of the expected high resistance of the interstitial space~\cite{Holter2018, vinje2020intracranial, tithof2021network, rey2018pulsatile}, we expect the perivascular-interstitial transfer and interstitial flow to be relatively small under physiological conditions. However, in light of the importance of quantifying and characterizing the different potential pathways, coupled fluid dynamics in vascular, perivascular and interstitial spaces will be considered in subsequent work.

We here consider open (in contrast to porous) domains. This is an appropriate modelling choice for surface perivascular spaces surrounding arteries or veins~\cite{min2020surface, bedussi2018paravascular}. For parenchymal perivascular spaces, within the pial-glial interface or within the smooth muscle cell basement membranes~\cite{albargothy2018convective}, however, a porous media representation may be more appropriate. In such a case, the Stokes flow equations~\eqref{eq:stokes} are naturally replaced by a Darcy or Brinkman flow model with an additional permeability $\kappa$~\cite{brinkman1949calculation}. The analogous reduced model (corresponding to~\eqref{eq:1d-pvs}) would include an additional lower order term for the flux $\hat{q}$ weighted by this permeability. For parenchymal and capillary perivascular spaces, we would also expect the coupled interplay between vascular, perivascular and interstitial spaces to be non-negligible. 

Furthermore, we have approximated the PVS as an (elliptic) annular structure, while surface PVSs may be of different shapes~\cite{tithof2019hydraulic, bedussi2018paravascular, vinje2021brain}. An interesting point is the quantification of the model error introduced by approximating these non-regular structures by elliptic annular cylinders with a fixed centerline. Gjerde et al \cite{gjerde2021analysis} addresses this point numerically and via theoretical analysis, including the balance between numerical and model errors. Finally, we also note that we have considered simplified (prescribed traction) boundary conditions at the PVS inlet and outlets. Compliance or resistance-based boundary conditions could of course also be considered, e.g.~as in previous work~\cite{daversin2020mechanisms}.  
We have focused on cardiac pulse wave-induced wall motion and vasomotion, two physiological factors that generate changes in vascular radius of up to $15\%$~\cite{mestre2018flow, aldea2019cerebrovascular} and only moderate wall velocities. However, the vascular and perivascular diameters may change more dramatically. For instance, Enger et al~\cite{enger2015dynamics} report of a nearly $40\%$ increase and $50\%$ decrease in arteriole diameter during cortical spreading depression, and intriguingly the vascular and perivascular wall motions may differ between e.g.~sleep states~\cite{bojarskaite2020astrocytic}. If these changes lead to significantly higher wall velocities than those considered here, we would expect a further breakdown of the reduced model assumptions, specifically assumption $\textbf{V}$, which in turn would be expected to impact the accuracy of the reduced models.

While many aspects of brain influx and clearance remain enigmatic, perivascular fluid flow along the cerebral vasculature is recognized as a key transport mechanism. The computationally inexpensive yet accurate reduced models presented here give an efficient and flexible framework for computational modelling and simulation of pulsatile flow in idealized or realistic networks including complete representations of e.g. the cerebral arteries or veins and many generations of arterioles/capillaries. This framework thus establishes a foundation for future computational studies of perivascular flow to improve our understanding of brain transport.  

\section*{Acknowledgements}

This study has received funding from the European Research Council (ERC) under the European Union's Horizon 2020 research and innovation programme under grant agreement 714892.  

\section*{Competing interests}
The authors declare that they have no competing interests.

\bibliography{references}

\begin{thebibliography}{10}
\urlstyle{rm}
\expandafter\ifx\csname url\endcsname\relax
  \def\url#1{\texttt{#1}}\fi
\expandafter\ifx\csname urlprefix\endcsname\relax\def\urlprefix{URL }\fi
\expandafter\ifx\csname doiprefix\endcsname\relax\def\doiprefix{DOI: }\fi
\providecommand{\bibinfo}[2]{#2}
\providecommand{\eprint}[2][]{\url{#2}}

\bibitem{rennels1985evidence}
\bibinfo{author}{Rennels, M.~L.}, \bibinfo{author}{Gregory, T.~F.},
  \bibinfo{author}{Blaumanis, O.~R.}, \bibinfo{author}{Fujimoto, K.} \&
  \bibinfo{author}{Grady, P.~A.}
\newblock \bibinfo{journal}{\bibinfo{title}{Evidence for a
  ‘paravascular’fluid circulation in the mammalian central nervous system,
  provided by the rapid distribution of tracer protein throughout the brain
  from the subarachnoid space}}.
\newblock {\emph{\JournalTitle{Brain research}}}
  \textbf{\bibinfo{volume}{326}}, \bibinfo{pages}{47--63}
  (\bibinfo{year}{1985}).

\bibitem{carare2008solutes}
\bibinfo{author}{Carare, R.} \emph{et~al.}
\newblock \bibinfo{journal}{\bibinfo{title}{Solutes, but not cells, drain from
  the brain parenchyma along basement membranes of capillaries and arteries:
  significance for cerebral amyloid angiopathy and neuroimmunology}}.
\newblock {\emph{\JournalTitle{Neuropathology and applied neurobiology}}}
  \textbf{\bibinfo{volume}{34}}, \bibinfo{pages}{131--144}
  (\bibinfo{year}{2008}).

\bibitem{iliff2012paravascular}
\bibinfo{author}{Iliff, J.~J.} \emph{et~al.}
\newblock \bibinfo{journal}{\bibinfo{title}{{A paravascular pathway facilitates
  CSF flow through the brain parenchyma and the clearance of interstitial
  solutes, including amyloid-$\beta$}}}.
\newblock {\emph{\JournalTitle{Science translational medicine}}}
  \textbf{\bibinfo{volume}{4}}, \bibinfo{pages}{147ra111--147ra111}
  (\bibinfo{year}{2012}).

\bibitem{wardlaw2020perivascular}
\bibinfo{author}{Wardlaw, J.~M.} \emph{et~al.}
\newblock \bibinfo{journal}{\bibinfo{title}{Perivascular spaces in the brain:
  anatomy, physiology and pathology}}.
\newblock {\emph{\JournalTitle{Nature Reviews Neurology}}}
  \textbf{\bibinfo{volume}{16}}, \bibinfo{pages}{137--153}
  (\bibinfo{year}{2020}).

\bibitem{zhang1990interrelationships}
\bibinfo{author}{Zhang, E.}, \bibinfo{author}{Inman, C.} \&
  \bibinfo{author}{Weller, R.}
\newblock \bibinfo{journal}{\bibinfo{title}{{Interrelationships of the pia
  mater and the perivascular (Virchow-Robin) spaces in the human cerebrum}}}.
\newblock {\emph{\JournalTitle{Journal of anatomy}}}
  \textbf{\bibinfo{volume}{170}}, \bibinfo{pages}{111} (\bibinfo{year}{1990}).

\bibitem{bedussi2018paravascular}
\bibinfo{author}{Bedussi, B.}, \bibinfo{author}{Almasian, M.},
  \bibinfo{author}{de~Vos, J.}, \bibinfo{author}{VanBavel, E.} \&
  \bibinfo{author}{Bakker, E.~N.}
\newblock \bibinfo{journal}{\bibinfo{title}{Paravascular spaces at the brain
  surface: Low resistance pathways for cerebrospinal fluid flow}}.
\newblock {\emph{\JournalTitle{Journal of Cerebral Blood Flow \& Metabolism}}}
  \textbf{\bibinfo{volume}{38}}, \bibinfo{pages}{719--726}
  (\bibinfo{year}{2018}).

\bibitem{tithof2019hydraulic}
\bibinfo{author}{Tithof, J.}, \bibinfo{author}{Kelley, D.~H.},
  \bibinfo{author}{Mestre, H.}, \bibinfo{author}{Nedergaard, M.} \&
  \bibinfo{author}{Thomas, J.~H.}
\newblock \bibinfo{journal}{\bibinfo{title}{Hydraulic resistance of
  periarterial spaces in the brain}}.
\newblock {\emph{\JournalTitle{Fluids and Barriers of the CNS}}}
  \textbf{\bibinfo{volume}{16}}, \bibinfo{pages}{1--13} (\bibinfo{year}{2019}).

\bibitem{min2020surface}
\bibinfo{author}{Min~Rivas, F.} \emph{et~al.}
\newblock \bibinfo{journal}{\bibinfo{title}{Surface periarterial spaces of the
  mouse brain are open, not porous}}.
\newblock {\emph{\JournalTitle{Journal of the Royal Society Interface}}}
  \textbf{\bibinfo{volume}{17}}, \bibinfo{pages}{20200593}
  (\bibinfo{year}{2020}).

\bibitem{martinac2019computational}
\bibinfo{author}{Martinac, A.~D.} \& \bibinfo{author}{Bilston, L.~E.}
\newblock \bibinfo{journal}{\bibinfo{title}{Computational modelling of fluid
  and solute transport in the brain}}.
\newblock {\emph{\JournalTitle{Biomechanics and modeling in mechanobiology}}}
  \textbf{\bibinfo{volume}{xx}}, \bibinfo{pages}{1--20} (\bibinfo{year}{2019}).

\bibitem{faghih2018bulk}
\bibinfo{author}{Faghih, M.~M.} \& \bibinfo{author}{Sharp, M.~K.}
\newblock \bibinfo{journal}{\bibinfo{title}{{Is bulk flow plausible in
  perivascular, paravascular and paravenous channels?}}}
\newblock {\emph{\JournalTitle{Fluids and Barriers of the CNS}}}
  \textbf{\bibinfo{volume}{15}}, \bibinfo{pages}{17} (\bibinfo{year}{2018}).

\bibitem{asgari2016glymphatic}
\bibinfo{author}{Asgari, M.}, \bibinfo{author}{De~Z{\'e}licourt, D.} \&
  \bibinfo{author}{Kurtcuoglu, V.}
\newblock \bibinfo{journal}{\bibinfo{title}{Glymphatic solute transport does
  not require bulk flow}}.
\newblock {\emph{\JournalTitle{Scientific reports}}}
  \textbf{\bibinfo{volume}{6}}, \bibinfo{pages}{1--11} (\bibinfo{year}{2016}).

\bibitem{diem2017arterial}
\bibinfo{author}{Diem, A.~K.} \emph{et~al.}
\newblock \bibinfo{journal}{\bibinfo{title}{Arterial pulsations cannot drive
  intramural periarterial drainage: significance for {A}$\beta$ drainage}}.
\newblock {\emph{\JournalTitle{Frontiers in neuroscience}}}
  \textbf{\bibinfo{volume}{11}}, \bibinfo{pages}{475} (\bibinfo{year}{2017}).

\bibitem{rey2018pulsatile}
\bibinfo{author}{Rey, J.} \& \bibinfo{author}{Sarntinoranont, M.}
\newblock \bibinfo{journal}{\bibinfo{title}{Pulsatile flow drivers in brain
  parenchyma and perivascular spaces: a resistance network model study}}.
\newblock {\emph{\JournalTitle{Fluids and Barriers of the CNS}}}
  \textbf{\bibinfo{volume}{15}}, \bibinfo{pages}{20} (\bibinfo{year}{2018}).

\bibitem{sharp2019dispersion}
\bibinfo{author}{Sharp, M.~K.}, \bibinfo{author}{Carare, R.~O.} \&
  \bibinfo{author}{Martin, B.~A.}
\newblock \bibinfo{journal}{\bibinfo{title}{Dispersion in porous media in
  oscillatory flow between flat plates: applications to intrathecal,
  periarterial and paraarterial solute transport in the central nervous
  system}}.
\newblock {\emph{\JournalTitle{Fluids and Barriers of the CNS}}}
  \textbf{\bibinfo{volume}{16}}, \bibinfo{pages}{13} (\bibinfo{year}{2019}).

\bibitem{lloyd2019effects}
\bibinfo{author}{Lloyd, R.~A.}, \bibinfo{author}{Stoodley, M.~A.},
  \bibinfo{author}{Fletcher, D.~F.} \& \bibinfo{author}{Bilston, L.~E.}
\newblock \bibinfo{journal}{\bibinfo{title}{The effects of variation in the
  arterial pulse waveform on perivascular flow}}.
\newblock {\emph{\JournalTitle{Journal of biomechanics}}}
  \textbf{\bibinfo{volume}{90}}, \bibinfo{pages}{65--70}
  (\bibinfo{year}{2019}).

\bibitem{kedarasetti2020arterial}
\bibinfo{author}{Kedarasetti, R.~T.}, \bibinfo{author}{Drew, P.~J.} \&
  \bibinfo{author}{Costanzo, F.}
\newblock \bibinfo{journal}{\bibinfo{title}{{Arterial pulsations drive
  oscillatory flow of CSF but not directional pumping}}}.
\newblock {\emph{\JournalTitle{Scientific reports}}}
  \textbf{\bibinfo{volume}{10}}, \bibinfo{pages}{1--12} (\bibinfo{year}{2020}).

\bibitem{kedarasetti2020functional}
\bibinfo{author}{Kedarasetti, R.~T.} \emph{et~al.}
\newblock \bibinfo{journal}{\bibinfo{title}{Functional hyperemia drives fluid
  exchange in the paravascular space}}.
\newblock {\emph{\JournalTitle{Fluids and Barriers of the CNS}}}
  \textbf{\bibinfo{volume}{17}}, \bibinfo{pages}{1--25} (\bibinfo{year}{2020}).

\bibitem{daversin2020mechanisms}
\bibinfo{author}{Daversin-Catty, C.}, \bibinfo{author}{Vinje, V.},
  \bibinfo{author}{Mardal, K.-A.} \& \bibinfo{author}{Rognes, M.~E.}
\newblock \bibinfo{journal}{\bibinfo{title}{The mechanisms behind perivascular
  fluid flow}}.
\newblock {\emph{\JournalTitle{{PLOS} {ONE}}}} \textbf{\bibinfo{volume}{15}},
  \bibinfo{pages}{e0244442}, \doiprefix\url{10.1371/journal.pone.0244442}
  (\bibinfo{year}{2020}).

\bibitem{olufsen1999structured}
\bibinfo{author}{Olufsen, M.~S.}
\newblock \bibinfo{journal}{\bibinfo{title}{Structured tree outflow condition
  for blood flow in larger systemic arteries}}.
\newblock {\emph{\JournalTitle{American journal of physiology-Heart and
  circulatory physiology}}} \textbf{\bibinfo{volume}{276}},
  \bibinfo{pages}{H257--H268} (\bibinfo{year}{1999}).

\bibitem{sherwin2003one}
\bibinfo{author}{Sherwin, S.}, \bibinfo{author}{Franke, V.},
  \bibinfo{author}{Peir{\'o}, J.} \& \bibinfo{author}{Parker, K.}
\newblock \bibinfo{journal}{\bibinfo{title}{One-dimensional modelling of a
  vascular network in space-time variables}}.
\newblock {\emph{\JournalTitle{Journal of engineering mathematics}}}
  \textbf{\bibinfo{volume}{47}}, \bibinfo{pages}{217--250}
  (\bibinfo{year}{2003}).

\bibitem{d2008coupling}
\bibinfo{author}{D'Angelo, C.} \& \bibinfo{author}{Quarteroni, A.}
\newblock \bibinfo{journal}{\bibinfo{title}{On the coupling of 1d and 3d
  diffusion-reaction equations: application to tissue perfusion problems}}.
\newblock {\emph{\JournalTitle{Mathematical Models and Methods in Applied
  Sciences}}} \textbf{\bibinfo{volume}{18}}, \bibinfo{pages}{1481--1504}
  (\bibinfo{year}{2008}).

\bibitem{lesinigo2011multiscale}
\bibinfo{author}{Lesinigo, M.}, \bibinfo{author}{D’Angelo, C.} \&
  \bibinfo{author}{Quarteroni, A.}
\newblock \bibinfo{journal}{\bibinfo{title}{{A multiscale Darcy--Brinkman model
  for fluid flow in fractured porous media}}}.
\newblock {\emph{\JournalTitle{Numerische Mathematik}}}
  \textbf{\bibinfo{volume}{117}}, \bibinfo{pages}{717--752}
  (\bibinfo{year}{2011}).

\bibitem{coccarelli2021framework}
\bibinfo{author}{Coccarelli, A.}, \bibinfo{author}{Carson, J.~M.},
  \bibinfo{author}{Aggarwal, A.} \& \bibinfo{author}{Pant, S.}
\newblock \bibinfo{journal}{\bibinfo{title}{A framework for incorporating 3d
  hyperelastic vascular wall models in 1d blood flow simulations}}.
\newblock {\emph{\JournalTitle{Biomechanics and Modeling in Mechanobiology}}}
  \bibinfo{pages}{1--19} (\bibinfo{year}{2021}).

\bibitem{Kppl2020}
\bibinfo{author}{K\"{o}ppl, T.}, \bibinfo{author}{Vidotto, E.} \&
  \bibinfo{author}{Wohlmuth, B.}
\newblock \bibinfo{journal}{\bibinfo{title}{A 3d-1d coupled blood flow and
  oxygen transport model to generate microvascular networks}}.
\newblock {\emph{\JournalTitle{International Journal for Numerical Methods in
  Biomedical Engineering}}} \textbf{\bibinfo{volume}{36}},
  \doiprefix\url{10.1002/cnm.3386} (\bibinfo{year}{2020}).

\bibitem{Koch2020}
\bibinfo{author}{Koch, T.}, \bibinfo{author}{Schneider, M.},
  \bibinfo{author}{Helmig, R.} \& \bibinfo{author}{Jenny, P.}
\newblock \bibinfo{journal}{\bibinfo{title}{Modeling tissue perfusion in terms
  of 1d-3d embedded mixed-dimension coupled problems with distributed
  sources}}.
\newblock {\emph{\JournalTitle{Journal of Computational Physics}}}
  \textbf{\bibinfo{volume}{410}}, \bibinfo{pages}{109370},
  \doiprefix\url{10.1016/j.jcp.2020.109370} (\bibinfo{year}{2020}).

\bibitem{vidotto2018}
\bibinfo{author}{Vidotto, E.}, \bibinfo{author}{Koch, T.},
  \bibinfo{author}{K\"{o}ppl, T.}, \bibinfo{author}{Helmig, R.} \&
  \bibinfo{author}{Wohlmuth, B.}
\newblock \bibinfo{journal}{\bibinfo{title}{Hybrid models for simulating blood
  flow in microvascular networks}}.
\newblock {\emph{\JournalTitle{Multiscale Modeling {\&} Simulation}}}
  \textbf{\bibinfo{volume}{17}}, \bibinfo{pages}{1076--1102},
  \doiprefix\url{10.1137/18m1228712} (\bibinfo{year}{2019}).

\bibitem{cattaneo2014}
\bibinfo{author}{Cattaneo, L.} \& \bibinfo{author}{Zunino, P.}
\newblock \bibinfo{journal}{\bibinfo{title}{A computational model of drug
  delivery through microcirculation to compare different tumor treatments}}.
\newblock {\emph{\JournalTitle{International Journal for Numerical Methods in
  Biomedical Engineering}}} \textbf{\bibinfo{volume}{30}},
  \bibinfo{pages}{1347--1371}, \doiprefix\url{10.1002/cnm.2661}
  (\bibinfo{year}{2014}).

\bibitem{Possenti2018}
\bibinfo{author}{Possenti, L.} \emph{et~al.}
\newblock \bibinfo{journal}{\bibinfo{title}{A computational model for
  microcirculation including fahraeus-lindqvist effect, plasma skimming and
  fluid exchange with the tissue interstitium}}.
\newblock {\emph{\JournalTitle{International Journal for Numerical Methods in
  Biomedical Engineering}}} \textbf{\bibinfo{volume}{35}},
  \bibinfo{pages}{e3165}, \doiprefix\url{10.1002/cnm.3165}
  (\bibinfo{year}{2018}).

\bibitem{Possenti2021}
\bibinfo{author}{Possenti, L.} \emph{et~al.}
\newblock \bibinfo{journal}{\bibinfo{title}{A mesoscale computational model for
  microvascular oxygen transfer}}.
\newblock {\emph{\JournalTitle{Annals of Biomedical Engineering}}}
  \doiprefix\url{10.1007/s10439-021-02807-x} (\bibinfo{year}{2021}).

\bibitem{mestre2018flow}
\bibinfo{author}{Mestre, H.} \emph{et~al.}
\newblock \bibinfo{journal}{\bibinfo{title}{Flow of cerebrospinal fluid is
  driven by arterial pulsations and is reduced in hypertension}}.
\newblock {\emph{\JournalTitle{Nature Communications}}}
  \textbf{\bibinfo{volume}{9}}, \doiprefix\url{10.1038/s41467-018-07318-3}
  (\bibinfo{year}{2018}).

\bibitem{pvs-meshing-tools}
\bibinfo{author}{Daversin-Catty, C.}
\newblock \bibinfo{title}{{PVS meshing tools}}.
\newblock \bibinfo{howpublished}{Github} (\bibinfo{year}{2020}).

\bibitem{VMTKAntiga2008AnIM}
\bibinfo{author}{Antiga, L.} \emph{et~al.}
\newblock \bibinfo{journal}{\bibinfo{title}{An image-based modeling framework
  for patient-specific computational hemodynamics}}.
\newblock {\emph{\JournalTitle{Medical {\&} Biological Engineering {\&}
  Computing}}} \textbf{\bibinfo{volume}{46}}, \bibinfo{pages}{1097--1112}
  (\bibinfo{year}{2008}).

\bibitem{meshio}
\bibinfo{author}{Schlömer, N.} \& \bibinfo{author}{al}.
\newblock \bibinfo{title}{meshio v4.3.10}.
\newblock \bibinfo{howpublished}{Zenodo} (\bibinfo{year}{2020}).

\bibitem{GMSH}
\bibinfo{author}{Geuzaine, C.} \& \bibinfo{author}{Remacle, J.-F.}
\newblock \bibinfo{journal}{\bibinfo{title}{Gmsh: A 3-d finite element mesh
  generator with built-in pre- and post-processing facilities}}.
\newblock {\emph{\JournalTitle{International Journal for Numerical Methods in
  Engineering}}} \textbf{\bibinfo{volume}{79}}, \bibinfo{pages}{1309--1331},
  \doiprefix\url{https://doi.org/10.1002/nme.2579} (\bibinfo{year}{2009}).

\bibitem{san2009convergence}
\bibinfo{author}{San~Mart{\'\i}n, J.}, \bibinfo{author}{Smaranda, L.} \&
  \bibinfo{author}{Takahashi, T.}
\newblock \bibinfo{journal}{\bibinfo{title}{Convergence of a finite
  element/{ALE} method for the {Stokes} equations in a domain depending on
  time}}.
\newblock {\emph{\JournalTitle{Journal of computational and applied
  mathematics}}} \textbf{\bibinfo{volume}{230}}, \bibinfo{pages}{521--545}
  (\bibinfo{year}{2009}).

\bibitem{aldea2019cerebrovascular}
\bibinfo{author}{Aldea, R.}, \bibinfo{author}{Weller, R.~O.},
  \bibinfo{author}{Wilcock, D.~M.}, \bibinfo{author}{Carare, R.~O.} \&
  \bibinfo{author}{Richardson, G.}
\newblock \bibinfo{journal}{\bibinfo{title}{Cerebrovascular smooth muscle cells
  as the drivers of intramural periarterial drainage of the brain}}.
\newblock {\emph{\JournalTitle{Frontiers in aging neuroscience}}}
  \textbf{\bibinfo{volume}{11}}, \bibinfo{pages}{1} (\bibinfo{year}{2019}).

\bibitem{gjerde2021analysis}
\bibinfo{author}{Gjerde, I.~G.}, \bibinfo{author}{Daversin-Catty, C.} \&
  \bibinfo{author}{Rognes, M.~E.}
\newblock \bibinfo{journal}{\bibinfo{title}{Analysis of one-dimensional flow
  models for perivascular fluid flow}}.
\newblock {\emph{\JournalTitle{In preparation}}}  (\bibinfo{year}{2021}).

\bibitem{AlnaesBlechta2015a}
\bibinfo{author}{Aln{\ae}s, M.~S.} \emph{et~al.}
\newblock \bibinfo{journal}{\bibinfo{title}{The {FEniCS Project Version 1.5}}}.
\newblock {\emph{\JournalTitle{Archive of Numerical Software}}}
  \textbf{\bibinfo{volume}{3}}, \bibinfo{pages}{9--23},
  \doiprefix\url{10.11588/ans.2015.100.20553} (\bibinfo{year}{2015}).

\bibitem{mechanisms-behind-pvs-flow-zenodo}
\bibinfo{author}{Daversin-Catty, C.}, \bibinfo{author}{Vinje, V.},
  \bibinfo{author}{Mardal, K.-A.} \& \bibinfo{author}{Rognes, M.~E.}
\newblock \bibinfo{title}{mechanisms-behind-pvs-flow-v1.0},
  \doiprefix\url{10.5281/zenodo.3890133} (\bibinfo{year}{2020}).

\bibitem{daversincatty2019abstractions}
\bibinfo{author}{Daversin-Catty, C.}, \bibinfo{author}{{Richardson}, C.~N.},
  \bibinfo{author}{{Ellingsrud}, A.~J.} \& \bibinfo{author}{{Rognes}, M.~E.}
\newblock \bibinfo{journal}{\bibinfo{title}{Abstractions and automated
  algorithms for mixed-dimensional finite element methods}}.
\newblock {\emph{\JournalTitle{ACM Transactions on Mathematical Software}}}
  (\bibinfo{year}{2021}).

\bibitem{Kuchta2020}
\bibinfo{author}{Kuchta, M.}
\newblock \bibinfo{title}{Assembly of multiscale linear {PDE} operators}.
\newblock In \emph{\bibinfo{booktitle}{Lecture Notes in Computational Science
  and Engineering}}, \bibinfo{pages}{641--650},
  \doiprefix\url{10.1007/978-3-030-55874-1_63} (\bibinfo{publisher}{Springer
  International Publishing}, \bibinfo{year}{2020}).

\bibitem{olufsen2004}
\bibinfo{author}{Olufsen, M.} \& \bibinfo{author}{Nadim, A.}
\newblock \bibinfo{journal}{\bibinfo{title}{On deriving lumped models for blood
  flow and pressure in the systemic arteries}}.
\newblock {\emph{\JournalTitle{Mathematical biosciences and engineering :
  MBE}}} \textbf{\bibinfo{volume}{1}}, \bibinfo{pages}{61--80},
  \doiprefix\url{10.3934/mbe.2004.1.61} (\bibinfo{year}{2004}).

\bibitem{tithof2021network}
\bibinfo{author}{Tithof, J.} \emph{et~al.}
\newblock \bibinfo{journal}{\bibinfo{title}{A network model of glymphatic flow
  under different experimentally-motivated parametric scenarios}}.
\newblock {\emph{\JournalTitle{bioRxiv}}}
  \doiprefix\url{10.1101/2021.09.23.461519} (\bibinfo{year}{2021}).

\bibitem{notaro2016mixed}
\bibinfo{author}{Notaro, D.}, \bibinfo{author}{Cattaneo, a.},
  \bibinfo{author}{Formaggia, L.}, \bibinfo{author}{Scotti, A.} \&
  \bibinfo{author}{Zunino, P.}
\newblock \bibinfo{title}{A mixed finite element method for modeling the fluid
  exchange between microcirculation and tissue interstitium}.
\newblock In \bibinfo{editor}{Ventura, G.} \& \bibinfo{editor}{Benvenuti, E.}
  (eds.) \emph{\bibinfo{booktitle}{Advances in Discretization Methods:
  Discontinuities, Virtual Elements, Fictitious Domain Methods}},
  \bibinfo{pages}{3--25}, \doiprefix\url{10.1007/978-3-319-41246-7_1}
  (\bibinfo{publisher}{Springer International Publishing},
  \bibinfo{address}{Cham}, \bibinfo{year}{2016}).

\bibitem{Holter2018}
\bibinfo{author}{Holter, K.~E.}, \bibinfo{author}{Kuchta, M.} \&
  \bibinfo{author}{Mardal, K.-A.}
\newblock \bibinfo{title}{Sub-voxel perfusion modeling in terms of coupled
  3d-1d problem}.
\newblock In \bibinfo{editor}{Radu, F.~A.}, \bibinfo{editor}{Kumar, K.},
  \bibinfo{editor}{Berre, I.}, \bibinfo{editor}{Nordbotten, J.~M.} \&
  \bibinfo{editor}{Pop, I.~S.} (eds.) \emph{\bibinfo{booktitle}{Numerical
  Mathematics and Advanced Applications ENUMATH 2017}}, \bibinfo{pages}{35--47}
  (\bibinfo{publisher}{Springer International Publishing},
  \bibinfo{address}{Cham}, \bibinfo{year}{2019}).

\bibitem{vinje2020intracranial}
\bibinfo{author}{Vinje, V.}, \bibinfo{author}{Eklund, A.},
  \bibinfo{author}{Mardal, K.-A.}, \bibinfo{author}{Rognes, M.~E.} \&
  \bibinfo{author}{St{\o}verud, K.-H.}
\newblock \bibinfo{journal}{\bibinfo{title}{{Intracranial pressure elevation
  alters CSF clearance pathways}}}.
\newblock {\emph{\JournalTitle{Fluids and Barriers of the CNS}}}
  \textbf{\bibinfo{volume}{17}}, \bibinfo{pages}{1--19} (\bibinfo{year}{2020}).

\bibitem{albargothy2018convective}
\bibinfo{author}{Albargothy, N.~J.} \emph{et~al.}
\newblock \bibinfo{journal}{\bibinfo{title}{{Convective influx/glymphatic
  system: tracers injected into the CSF enter and leave the brain along
  separate periarterial basement membrane pathways}}}.
\newblock {\emph{\JournalTitle{Acta neuropathologica}}}
  \textbf{\bibinfo{volume}{136}}, \bibinfo{pages}{139--152}
  (\bibinfo{year}{2018}).

\bibitem{brinkman1949calculation}
\bibinfo{author}{Brinkman, H.}
\newblock \bibinfo{journal}{\bibinfo{title}{A calculation of the viscous force
  exerted by a flowing fluid on a dense swarm of particles}}.
\newblock {\emph{\JournalTitle{Flow, Turbulence and Combustion}}}
  \textbf{\bibinfo{volume}{1}}, \bibinfo{pages}{27--34} (\bibinfo{year}{1949}).

\bibitem{vinje2021brain}
\bibinfo{author}{Vinje, V.}, \bibinfo{author}{Bakker, E. N.~T.~P.} \&
  \bibinfo{author}{Rognes, M.~E.}
\newblock \bibinfo{journal}{\bibinfo{title}{Brain solute transport is more
  rapid in periarterial than perivenous spaces}}.
\newblock {\emph{\JournalTitle{Scientific Reports}}}  (\bibinfo{year}{2021}).

\bibitem{enger2015dynamics}
\bibinfo{author}{Enger, R.} \emph{et~al.}
\newblock \bibinfo{journal}{\bibinfo{title}{Dynamics of ionic shifts in
  cortical spreading depression}}.
\newblock {\emph{\JournalTitle{Cerebral Cortex}}}
  \textbf{\bibinfo{volume}{25}}, \bibinfo{pages}{4469--4476}
  (\bibinfo{year}{2015}).

\bibitem{bojarskaite2020astrocytic}
\bibinfo{author}{Bojarskaite, L.} \emph{et~al.}
\newblock \bibinfo{journal}{\bibinfo{title}{Astrocytic ca 2+ signaling is
  reduced during sleep and is involved in the regulation of slow wave sleep}}.
\newblock {\emph{\JournalTitle{Nature communications}}}
  \textbf{\bibinfo{volume}{11}}, \bibinfo{pages}{1--16} (\bibinfo{year}{2020}).

\end{thebibliography}

\end{document}